\documentclass[12pt,preprint]{elsarticle}
\usepackage{hyperref,lineno,natbib,amsmath,amssymb,graphicx,multirow,algorithmic,color,microtype,fullpage}
\usepackage[skip=-2pt,font=footnotesize]{caption}
\pdfoutput=1

\usepackage{mathptmx,textcomp}

\journal{Journal of Computational Physics}
\definecolor{dblue}{rgb}{0.0,0.0,0.5}
\definecolor{dmag}{rgb}{0.831,0.165,1.0}
\definecolor{dred}{rgb}{1.0,0,0}
\definecolor{jade}{rgb}{0.1333,0.5647,0.4784}
\definecolor{lblue}{rgb}{0,0.6745,1.0}
\definecolor{pmag}{rgb}{0.580,0.129,0.572}
\definecolor{pgry}{rgb}{0.572,0.572,0.572}

\usepackage{hyperref}

\definecolor{webgreen}{rgb}{0,.35,0}
\definecolor{webbrown}{rgb}{.6,0,0}
\definecolor{RoyalBlue}{rgb}{0,0,0.9}
\definecolor{mywhite}{rgb}{1.0,1.0,1.0}
\newcommand{\white}[1]{\textcolor{mywhite}{#1}}

\hypersetup{
   colorlinks=true, linktocpage=true, pdfstartpage=3, pdfstartview=FitV,
   breaklinks=true, pdfpagemode=UseNone, pageanchor=true, pdfpagemode=UseOutlines,
   plainpages=false, bookmarksnumbered, bookmarksopen=true, bookmarksopenlevel=1,
   hypertexnames=true, pdfhighlight=/O,
   urlcolor=webbrown, linkcolor=RoyalBlue, citecolor=webgreen,
   pdfauthor={Chris H. Rycroft, Yi Sui, and Eran Bouchbinder},
   pdfsubject={An Eulerian projection method for quasi-static elastoplasticity},
   pdfkeywords={fluid mechanics, projection method, plasticity, elastoplasticity},
   pdfcreator={pdfLaTeX},
   pdfproducer={LaTeX with hyperref}
}

\newcommand{\mcD}{\mathcal{D}}

\newcommand{\p}{\partial}
\renewcommand{\vec}[1]{\mathbf{#1}}
\newcommand{\ten}[1]{\mathbf{#1}}
\newcommand{\tC}{\ten{C}}
\newcommand{\tS}{\ten{S}}
\newcommand{\vx}{\vec{x}}
\newcommand{\vv}{\vec{v}}
\newcommand{\prx}[1]{\frac{\p #1}{\p x}}
\newcommand{\pry}[1]{\frac{\p #1}{\p y}}
\newcommand{\drt}[1]{\frac{d #1}{d t}}
\newcommand{\tD}{\ten{D}}
\newcommand{\Del}{\tD^\text{el}}
\newcommand{\Dpl}{\tD^\text{pl}}
\newcommand{\dpl}{D^\text{pl}}
\newcommand{\tdpln}{\tilde{D}^\text{pl}_n}
\newcommand{\Dt}{\Delta t}
\newcommand{\bsig}{\boldsymbol\sigma}
\newcommand{\bs}{\bar{s}}
\newcommand{\sC}{\mathcal{C}}
\newcommand{\sR}{\mathcal{R}}

\newcommand{\nor}{\hat{\vec{n}}}
\newcommand{\nvT}{\widetilde{\nabla\vv}}
\newcommand{\bome}{\boldsymbol\omega}
\newcommand{\tT}{\tilde{t}}

\newcommand{\sY}{s_\text{Y}}
\newcommand{\vcspv}{V_\vv}
\newcommand{\vcsps}{V_{\bsig}}
\newcommand{\visc}{\kappa}
\newcommand{\adv}{(\vv_n\cdot\nabla)}
\newcommand{\bxi}{\boldsymbol\xi}
\newcommand{\afac}{\eta}
\newcommand{\Dsig}{\bsig_\text{P}}

\newcommand{\tsca}{t_s}
\newcommand{\trel}{t_\textrm{R}}
\newcommand{\ted}{\text{E}}
\newcommand{\teq}{\text{Q}}
\newcommand{\hf}{\frac{1}{2}}
\newcommand{\thf}{\frac{3}{2}}
\newcommand{\transpose}{\mathsf{T}}

\DeclareMathOperator{\tr}{tr}

\begin{document}
\begin{frontmatter}
\title{An Eulerian projection method for quasi-static elastoplasticity}
\author[SEAS,LBL]{Chris H. Rycroft\corref{cor1}}
\ead{chr@seas.harvard.edu}
\author[SFU]{Yi Sui}
\author[Weizmann]{Eran Bouchbinder}
\ead{eran.bouchbinder@weizmann.ac.il}
\cortext[cor1]{Corresponding author.}
\address[SEAS]{Paulson School of Engineering and Applied Sciences, Harvard University, Cambridge, MA 02138, United States}
\address[LBL]{Department of Mathematics, Lawrence Berkeley Laboratory,
Berkeley, CA 94720, United States}
\address[SFU]{Department of Mathematics, Simon Fraser University, Burnaby, British Columbia, V5A 1S6, Canada}
\address[Weizmann]{Chemical Physics Department, Weizmann Institute of Science, Rehovot 76100, Israel}

\begin{abstract}
  A well-established numerical approach to solve the Navier--Stokes equations
  for incompressible fluids is Chorin's projection method~\cite{chorin68},
  whereby the fluid velocity is explicitly updated, and then an elliptic
  problem for the pressure is solved, which is used to orthogonally project the
  velocity field to maintain the incompressibility constraint. In this paper,
  we develop a mathematical correspondence between Newtonian fluids in the
  incompressible limit and hypo-elastoplastic solids in the slow, quasi-static
  limit. Using this correspondence, we formulate a new fixed-grid, Eulerian
  numerical method for simulating quasi-static hypo-elastoplastic solids,
  whereby the stress is explicitly updated, and then an elliptic problem for
  the velocity is solved, which is used to orthogonally project the stress to
  maintain the quasi-staticity constraint. We develop a finite-difference
  implementation of the method and apply it to an elasto-viscoplastic model of
  a bulk metallic glass based on the shear transformation zone theory. We show
  that in a two-dimensional plane strain simple shear simulation, the method is
  in quantitative agreement with an explicit method. Like the fluid projection
  method, it is efficient and numerically robust, making it practical for a
  wide variety of applications. We also demonstrate that the method can be
  extended to simulate objects with evolving boundaries. We highlight a number
  of correspondences between incompressible fluid mechanics and quasi-static
  elastoplasticity, creating possibilities for translating other numerical
  methods between the two classes of physical problems.
\end{abstract}
\end{frontmatter}

\section{Introduction}
A wide variety of materials of scientific and technological importance exhibit
elastoplastic behavior, such as metals~\cite{lipinski89,champion03}, granular
materials~\cite{henann13}, aerogels~\cite{moner-girona99}, and amorphous solids
such as bulk metallic glasses (BMGs)~\cite{vaidyanathan01}. At low levels of
stress these materials typically behave elastically, so that the deformation
they undergo is reversible when the stress is removed. However, at higher
levels of stress, the material will start to yield, and undergo plastic,
irreversible deformation that will remain after the stress is removed.
Describing elastoplastic\footnote{Throughout this article, we use
``elastoplastic'' to refer to any material response that is a combination of
reversible elastic deformation and irreversible plastic deformation. This
includes, for example, rate-independent elastic--perfectly plastic models and
rate-dependent elasto-viscoplastic models.} behavior within a consistent
theoretical framework has been the subject of major research effort over many
decades, particularly from the 1950's onward. As described in a recent review
article~\cite{xiao06}, accurately characterizing elastoplastic behavior has
proved challenging, since it is not obvious how to separate the elastic and
plastic response at the microscopic level. Several different frameworks have
emerged, each of which is based on different assumptions of how the elastic and
plastic behavior are combined.

Currently, perhaps the most widely used framework to study elastoplastic
materials is hyper-elastoplasticity~\cite{lubliner08,gurtin10}. This model is
based on introducing an initial undeformed reference configuration of a
material. A time-dependent mapping is then employed, transforming the reference
configuration into the deformed configuration at a later time. The deformation
gradient tensor $\ten{F}$ is then defined as the Jacobian matrix of the
mapping, and represents how an infinitesimal material element is transformed. A
purely elastic material can then be described in terms of a constitutive law
that gives stress as a function of $\ten{F}$. To generalize this to
elastoplastic behavior, the Kr\"oner--Lee decomposition was developed, whereby
the deformation gradient tensor is viewed as the product of elastic and plastic
parts, $\ten{F}=\ten{F}_e \ten{F}_p$~\cite{kroner59,lee69}. This decomposition
has been successfully used to model the elastoplastic behavior of a variety of
materials such as metals and metallic glasses~\cite{anand05,su06,thamburaja07},
and can be carried out in commercial solid mechanics software such as
\textsc{Abaqus}. However, the decomposition has also been extensively debated
within the literature. For materials that undergo very large plastic
deformation and rearrangement, the notion of a mapping from an initial
configuration may become problematic. The decomposition is non-unique, whereby
the stress remains invariant under the transformation of the intermediate
configuration \smash{$(\ten{F}_e,\ten{F}_p)\mapsto(\ten{F}_e
\ten{R}^\transpose, \ten{R} \ten{F}_p)$} for an arbitrary rotation $\ten{R}$.
While $\ten{F}_e$ and $\ten{F}_p$ remain useful mathematical quantities, they
may no longer retain their expected physical interpretations~\cite{xiao06},
which has led to recent efforts to clarify this from a micromechanical
perspective, at least for crystalline solids~\cite{reina14}.

An alternative framework is hypo-elastoplasticity, which is based on an
additive decomposition of the Eulerian rate-of-deformation tensor into elastic
and plastic parts, \smash{$\tD=\Del+\Dpl$}~\cite{truesdell55,hill58,prager60}.
This approach has some drawbacks: it has mainly been applied to elastoplastic
simulations involving only linear elastic deformation, since it is difficult to
capture a nonlinear elastic strain response purely through $\Del$. In
particular, several researchers have noted some undesirable effects of the
decomposition~\cite{nagtegaal81,dienes79}, such as leaving a residual stress
after an elastic strain cycle~\cite{kojic87}. Furthermore, because the
framework is based on velocity as opposed to deformation, it can lead to the
build-up of numerical errors during
time-integration~\cite{eterovic90,hughes80}. However, because it is based on
Eulerian quantities, it does not depend on an undeformed configuration, which
is a potential advantage for materials undergoing large strains. The
aforementioned difficulties are typically minor in the limit of small elastic
deformation, and hence it may provide a reasonable framework for many materials
such as metals and metallic glasses that have large elastic constants.

Another feature of hypo-elastoplasticity is that it naturally fits within an
Eulerian, fixed-grid framework, and there are several recent trends in
numerical computation that make fixed-grid methods desirable. A fixed grid has
simpler topology, making it easier and more efficient to program, and simpler
to parallelize. Eulerian methods are also a natural environment in which
fluid--structure interactions are accounted for, since fixed-grid frameworks
are often the technique of choice for fluids~\cite{tannehill97,yu03}. Several
approaches for dealing with nonlinear hyperelasticity have been proposed by
treating the deformation gradient tensor as an Eulerian
field~\cite{liu01,plohr88,trangenstein91} or by introducing a reference map
field that describes the deformation from the initial undeformed
state~\cite{kamrin_thesis,cottet08,maitre09,kamrin12}. Other physical effects
such as coupling to electrical fields~\cite{chen14a} or the diffusion of
temperature fit well within an Eulerian framework. Some manufacturing processes
featuring continuous motion of material, such as
extrusion~\cite{zienkiewicz74}, are also well-suited to the Eulerian viewpoint.

Starting from the additive decomposition of $\tD$, and coupling it with a
continuum version of Newton's second law, one ends up with a closed system of
partial differential equations for velocity, stress, and typically a set of
additional internal variables. From this system a direct, explicit numerical
scheme can be constructed. The scheme resolves elastic waves in the material,
leading to a restriction on the numerical timestep due to the
Courant--Friedrichs--Lewy (CFL) condition. For many materials of interest, such
as metals, the elastic wave speed is on the order of kilometers per second,
which makes it prohibitive to simulate processes on physically relevant time
scales of seconds, hours, or days. Because of this, most applications of
hypo-elastoplasticity have been interested in rapid processes such as
impact~\cite{tran04}, or have scaled the elastic constants to be artificially
soft~\cite{rycroft12}. If one scales the hypo-elastoplasticity equations to
examine the long timescale and small velocity limit, one finds that the
continuum version of Newton's second law can be replaced with a constraint that
the stresses remain in quasi-static equilibrium.

In this paper, we show that there is a strong mathematical connection between
quasi-static hypo-elastoplasticity and the incompressible Navier--Stokes
equations. For an incompressible fluid, the relevant variables are the velocity
and pressure. There is an explicit update equation for velocity, and the
incompressibility constraint requires that the velocity remain divergence-free.
In this situation, a well-established method of solution is the projection
method of Chorin~\cite{chorin68}, described in detail in
Subsec.~\ref{sub:fproj}, whereby the fluid velocity is explicitly updated, and
then an elliptic problem for the pressure is solved, which is used to
orthogonally project the velocity field to maintain the incompressibility
constraint. By exploiting the mathematical correspondence, we have developed a
new numerical method for quasi-static elastoplasticity that is analogous to the
projection method for incompressible fluid dynamics. It takes an analogous
approach, whereby the stress is explicitly updated, and then an elliptic
problem for the velocity is solved, which is used to orthogonally project the
stress to maintain the quasi-staticity constraint.

To the best of our knowledge, this mathematical correspondence has not been
noted and explored in detail before, and the resultant numerical method based
on a projection step to restore quasi-staticity is distinct from existing
computational approaches. Some of the most well-established numerical methods
make use of an updated Lagrangian formulation and a mesh that deforms with the
material~\cite{hibbitt70,mcmeeking75,needleman85}. Ponthot~\cite{ponthot02}
developed an implicit simulation approach for elastoplasticity, although it
again makes use of a moving-mesh framework, leading to different mathematical
considerations. A number of authors developed and analyzed two-step algorithms
for rate-independent plasticity, which involve an elastic predictor step
followed by a plastic corrector step whereby the stress is projected to the
yield surface~\cite{wilkins63,ortiz83,ortiz85,simo86,bruhns94,simo98}. However,
this notion of a projection, which is carried out for each material element, is
distinctly different from the global stress projection that we develop here.

In Section~\ref{sec:methods}, we describe the mathematical correspondence to
incompressible fluid mechanics and the associated numerical procedure. In
Section~\ref{sec:numerics}, to illustrate the method, we develop a
finite-difference implementation of it to study a specific rate-dependent,
elasto-viscoplastic model of a bulk metallic glass based on the shear
transformation zone (STZ) theory. Originally developed by Falk and
Langer~\cite{falk98}, this model has undergone substantial
development~\cite{bouchbinder07,bouchbinder09}, and has been applied to a wide
variety of amorphous materials. The STZ model of the bulk metallic glass is an
appropriate numerical example, since BMGs can undergo large amounts of plastic
deformation in certain situations (such as at high temperature), and have
elastic moduli on the order of 10--100~GPa, meaning that experimental tests are
often in the quasi-static regime. A previous study that examined cavitation as
a fracture mechanism in the STZ model specifically described the long timescale
limit and made use of the quasi-staticity constraint for theoretical
analysis~\cite{bouchbinder08b}.

While our numerical examples focus on the STZ model of a BMG, we note that the
core of the numerical approach can be applied to a wide variety of plasticity
models and physical problems. It could apply to other descriptions of BMGs,
such as free-volume-based models~\cite{spaepen77,huang02,gao07}, which result
in equations with a similar mathematical structure. It could also be applied to
hypo-elastic materials or to rate-independent plasticity models. The method is
not limited to the finite-difference method, and alternative discretization
procedures could be used, such as the finite-volume or discontinuous Galerkin
methods.

The first numerical example we present is a BMG undergoing simple shear
deformation in a two-dimensional, plane strain, periodic geometry, which is
simple enough to allow for quantitative analysis (Section~\ref{sec:shearing}).
By choosing parameters appropriately, we quantitatively compare the
quasi-static projection method to the explicit scheme. We provide numerical
evidence that the two methods agree in the quasi-static limit. We also show
that the quasi-static method can simulate elastoplastic dynamics on physically
realistic timescales.

Many important problems of interest involve moving boundaries and hence we need
an Eulerian description of such evolving boundaries. In Section~\ref{sec:free}
we extend the method to implement a traction-free boundary condition at a
boundary described by the level set method~\cite{osher88,sethian,osher}.
Finally, since the projection method makes use of the same numerical framework
as the explicit scheme, the two methods can be interchanged making it possible
to simulate phenomena on multiple disparate timescales. We previously
demonstrated this capability to examine dynamic crack
propagation~\cite{rycroft12b}. Here, we present another case, of a bar that is
loaded on a slow, quasi-static timescale and then released, undergoing rapid
vibrations.

While many computational methods for elastoplasticity are already available, we
find that the numerical method developed here offers a useful practical
approach for dealing with hypo-elastoplastic materials in the quasi-static
limit. One of the main advantages of the fluid projection method is that it
maintains the incompressibility condition through a single algebraic problem
for the pressure, which is generally well-conditioned and can be carried out
efficiently, and we find that many of the same benefits remain valid for the
elasto-plasticity method we develop. Throughout the paper, we find a surprising
number of correspondences between the two methods, such as analogous
considerations for boundary conditions or the uniqueness of solutions. The
mathematical connection opens up interesting possibilities for translating
numerical methods for incompressible fluid mechanics over to quasi-static
elastoplasticity and \textit{vice versa}.

\section{Theoretical development}
\label{sec:methods}
\subsection{An elastoplastic material model}
We consider an elastoplastic material with velocity $\vv(\vx,t)$ and Cauchy
stress tensor $\bsig(\vx,t)$. The spin is defined as $\bome = (\nabla \vv - (\nabla
\vv)^\transpose)/2$, and the rate-of-deformation tensor is $\tD=(\nabla \vv + (\nabla
\vv)^\transpose)/2$. For an arbitrary field $f(\vx,t)$, we define the advective
derivative as $df/dt = \p f/\p t + (\vv\cdot \nabla) f$. Using the
hypo-elastoplastic kinematic relation, the rate-of-deformation tensor is
assumed to be the sum of elastic and plastic parts such that $\tD=\Del + \Dpl$.
The linear elastic constitutive relation is
\begin{equation}
  \frac{\mcD \bsig}{\mcD t} = \tC : \Del = \tC : (\tD - \Dpl),
  \label{eq:constit}
\end{equation}
where $\tC$ is a fourth-rank stiffness tensor, which for simplicity of
presentation is assumed to be isotropic, and constant in space and time. The
left hand side of Eq.~\ref{eq:constit} is the Jaumann objective stress rate,
$\mcD\bsig/\mcD t=d\bsig/dt + \bsig\cdot \bome - \bome\cdot \bsig$, which gives
the time-evolution of the stress taking into account translation and rotation
of the material, under the assumption that the elastic deformation is
small~\cite{bazant71}. By considering force balance, the velocity satisfies
\begin{equation}
  \rho \frac{d\vv}{dt} = \nabla \cdot \bsig,
  \label{eq:force}
\end{equation}
where $\rho$ is the density of the material. Taken together,
Eqs.~\ref{eq:constit} and \ref{eq:force} form a hyperbolic system of equations
from which a finite-difference simulation of an elastoplastic material can be
constructed. However, the hyperbolic system will resolve the propagation of
elastic waves, and therefore the timestep $\Delta t$ and grid spacing $\Delta
x$ must be chosen to satisfy the CFL condition for numerical stability to be
maintained. If $c_e$ is an elastic wave speed, then the timestep must satisfy
$\Delta t \le \Delta x/c_e$. For many problems of practical importance, such as
simulating metals, this will pose a prohibitively strong restriction. A typical
elastic wave speed would be on the order of kilometers per second, while a grid
spacing could be on the order of millimeters to micrometers, thus requiring a
timestep on the order of microseconds or smaller. This restriction would make
it infeasible to simulate real problems on the timescale of seconds, minutes,
or hours.

We now consider the limit when the deformation of the material happens on a
time scale that is much longer than the time for elastic waves to propagate
across the system. We rescale the equations in the limit of long times and
corresponding small velocity gradients by introducing
\begin{equation}
  \nabla\vv=\epsilon \nvT, \qquad t = \frac{\tT}{\epsilon},
  \label{eq:long_time_scale}
\end{equation}
where $\epsilon$ is a small dimensionless parameter. Under these scalings, the
constitutive equation becomes
\begin{equation}
  \label{eq:scaconst}
  \frac{\mcD \bsig}{\mcD \tT} = \tC : \left(\tilde{\tD} - \frac{\Dpl}{\epsilon} \right),
\end{equation}
where \smash{$\tilde{\tD}=(\nvT + (\nvT)^\transpose)/2$}, and the force balance
equation becomes
\begin{equation}
  \epsilon \rho \frac{d\vv}{d\tT} = \nabla \cdot \bsig.
\end{equation}
There are two occurrences of $\epsilon$ in these equations. The $\epsilon^{-1}$
in Eq.~\ref{eq:scaconst} signifies that over long durations, plastic
deformation will become increasingly important, while the $\epsilon$ term on
$d\vv/dt$ signifies that accelerations decrease in importance. Through these
considerations, one can approximate the material response by neglecting the
$d\vv/d\tT$ term to give
\begin{equation}
  \nabla \cdot \bsig = \vec{0},
  \label{eq:quasis}
\end{equation}
which physically states that forces remain in quasi-static equilibrium. A
numerical scheme could then be constructed using the constitutive equation
Eq.~\ref{eq:constit} subject to the constraint in Eq.~\ref{eq:quasis}. However,
this raises several questions. It is not clear how to update the velocity,
since the ability to explicitly time-integrate it is lost. It is also not clear
whether solutions of this system will match the solutions of the original
system.

\subsection{Review of the projection method for the incompressible Navier--Stokes equations}
\label{sub:fproj}
To make progress with the above problem, we now consider a different class of
problems involving an incompressible fluid with velocity $\vv$, pressure $p$,
and density $\rho$. The fluid velocity satisfies the Navier--Stokes equations,
\begin{equation}
  \rho \frac{d\vv}{dt} = - \nabla p + \nabla \cdot \ten{T},
  \label{eq:navsto}
\end{equation}
where $\ten{T}$ is the fluid stress tensor, and the fluid density evolves
according to
\begin{equation}
  \frac{d \rho}{dt} = - \rho(\nabla \cdot \vv).
  \label{eq:flcont}
\end{equation}
In addition, an equation of state linking the fluid density to the pressure
must be satisfied. For typical weakly compressible fluids, the equation $\rho -
\rho_0 = (p-p_0)/c^2$ is appropriate, where $\rho_0$ and $p_0$ are reference
densities and pressures respectively, and $c$ is a large constant that
corresponds to a sound wave speed through the fluid.

In a similar manner to the elastoplastic system of equations considered in the
previous section, Eqs.~\ref{eq:navsto} and \ref{eq:flcont} form a hyperbolic
system of equations that could be used to construct an explicit
finite-difference simulation of the fluid, but due to the CFL condition, the
presence of the sound speed places a severe restriction on the timestep size.
Again, for many practical problems, one may wish to consider time scales that
are much longer than the time for compressive waves to propagate across the
system. Looking at long times by introducing $t=\tT/\epsilon$ as in
Eq.~\ref{eq:long_time_scale}, one finds that
\begin{equation}
  \epsilon \frac{d \rho}{d\tT} =- \rho (\nabla \cdot \vv)
  \label{eq:flscale}
\end{equation}
which can be approximated by
\begin{equation}
  \nabla \cdot \vv=0
  \label{eq:flincomp}
\end{equation}
so that the velocity is divergence-free. The resultant system given by
Eqs.~\ref{eq:navsto} and \ref{eq:flincomp} are the incompressible
Navier--Stokes equations.

Numerical methods to simulate the incompressible Navier--Stokes equations have
been extensively studied and developed. In work by Chorin~\cite{chorin67},
aiming at addressing the constraint imposed by Eq.~\ref{eq:flincomp}, the
incompressible Navier--Stokes equations were simulated by examining the
compressible system as the parameter $c$ becomes large. Numerical evidence
shows that in the limit in which $c$ becomes large, the compressible solutions
approach the incompressible ones. This can also be understood by
introducing a vector space $\vcspv$ of all velocity fields. The divergence-free
solutions $\vv \in \vcspv$, which satisfy $\nabla \cdot \vv=0$, form a subspace
in $\vcspv$. In the compressible case, the $d\rho/dt$ term in
Eq.~\ref{eq:flscale}, in tandem with the pressure gradient in
Eq.~\ref{eq:navsto}, force the system toward being divergence-free.

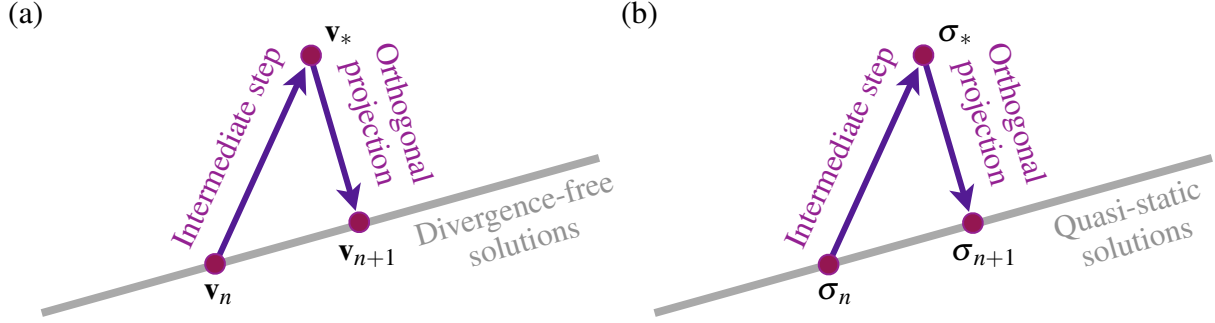
\begin{figure}
  \setlength{\unitlength}{0.745bp}
  \begin{center}
    {\normalsize
    \begin{picture}(620,175)(0,0)
      \put(0,0){\includegraphics[scale=0.745]{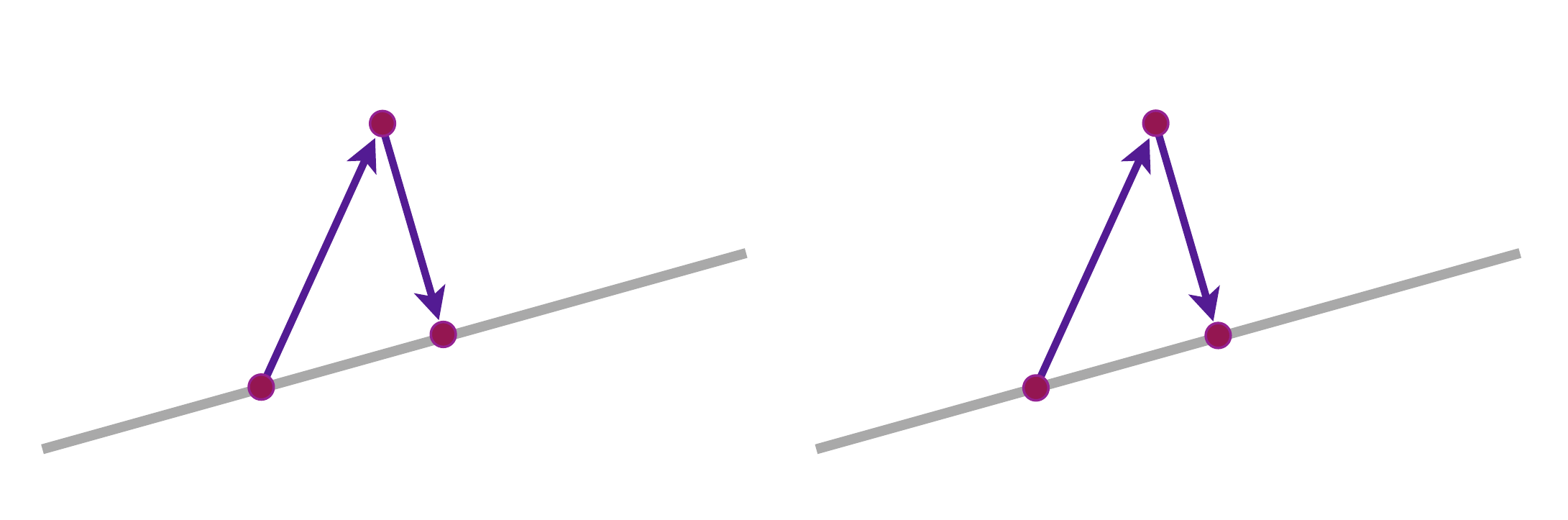}}
      \put(0,175){(a)}
      \put(99,35){$\vv_n$}
      \put(167,55){$\vv_{n+1}$}
      \put(160,166){$\vv_*$}
      \put(82,60){\rotatebox{65}{\textcolor{pmag}{Intermediate step}}}
      \put(149,217){\rotatebox{286}{\textcolor{pmag}{\parbox{5cm}{\centering Orthogonal\\\vspace{-0.5mm}projection}}}}
      \put(163,43){\rotatebox{16}{\textcolor{pgry}{\parbox{5cm}{\centering Divergence-free\\\vspace{-0.5mm}solutions}}}}
      \put(310,175){(b)}
      \put(409,35){$\bsig_n$}
      \put(477,55){$\bsig_{n+1}$}
      \put(470,166){$\bsig_*$}
      \put(392,60){\rotatebox{65}{\textcolor{pmag}{Intermediate step}}}
      \put(459,217){\rotatebox{286}{\textcolor{pmag}{\parbox{5cm}{\centering Orthogonal\\\vspace{-0.5mm}projection}}}}
      \put(473,43){\rotatebox{16}{\textcolor{pgry}{\parbox{5cm}{\centering Quasi-static\\\vspace{-0.5mm}solutions}}}}
    \end{picture}}
  \end{center}
  \caption{A schematic representation of the timestep in (a) the projection
  method for the incompressible Navier--Stokes equations and (b) the projection
  method for quasi-static elastoplasticity.\label{fig:schem}}
\end{figure}

This observation can be used as the basis of the projection method for
incompressible Navier--Stokes equations~\cite{chorin68}. Suppose that $\vv_n$
represents the discretized velocity field after $n$ steps in a
finite-difference simulation. To advance forward by $\Dt$ to the $(n+1)$th step
an intermediate velocity $\vv_*$ is first computed by neglecting the pressure
term, so that
\begin{equation}
  \frac{\rho(\vv_* - \vv_n)}{\Dt} =  - (\vv_n \cdot \nabla) \vv_n + \nabla \cdot \ten{T}_n.
  \label{eq:fproj1}
\end{equation}
If the pressure at the $(n+1)$th step was known then $\vv_{n+1}$ could be computed
according to
\begin{equation}
  \frac{\vv_{n+1} - \vv_*}{\Dt} = - \frac{1}{\rho} \nabla p_{n+1}.
  \label{eq:fproj2}
\end{equation}
Taking the divergence of Eq.~\ref{eq:fproj2} and enforcing that $\nabla \cdot \vv_{n+1}=0$ gives
\begin{equation}
  \nabla \cdot \vv_* = \frac{\Dt}{\rho} \nabla \cdot (\nabla p_{n+1}) = \frac{\Dt}{\rho} \nabla^2 p_{n+1}
  \label{eq:fproj3}
\end{equation}
and hence the pressure satisfies a Poisson equation where the source term is
$\nabla \cdot \vv_*$, which is an elliptic problem that can be solved
numerically using linear algebra. Boundary conditions on $p$ in this elliptic
problem depend on the specific situation considered, with the two most common
being a Dirichlet condition for a constant pressure boundary condition, or a
Neumann condition arising from a condition on the normal velocity component.
Once $p_{n+1}$ is evaluated, Eq.~\ref{eq:fproj2} can then be used to calculate
$\vv_{n+1}$. A schematic representation of the method in the vector space
$\vcspv$ is shown in Fig.~\ref{fig:schem}(a). The intermediate velocity may not
be in the divergence-free subspace, but the combination of Eqs.~\ref{eq:fproj2}
\& \ref{eq:fproj3} ensures that it is projected back to this subspace.

For consistency, it is also necessary to show that the projection applied by
Eq.~\ref{eq:fproj2} is in some sense orthogonal to the divergence-free
subspace. To do this, $\vcspv$ can be endowed with an inner product, where
for any $\vec{a},\vec{b} \in \vcspv$,
\begin{equation}
  \langle \vec{a},\vec{b} \rangle = \int \vec{a} \cdot \vec{b} \, d^3\vx.
  \label{eq:finprod}
\end{equation}
Hence, if problem-specific boundary terms are neglected, the projection
$\vv_\text{P}=\vv_{n+1}-\vv_*$ satisfies
\begin{equation}
  \langle \vv_{n+1} - \vv_n , \vv_\text{P} \rangle = -\frac{\Delta t}{\rho} \int (\vv_{n+1} - \vv_n) \cdot \nabla p_{n+1} \, d^3\vx = \frac{\Delta t}{\rho} \int (\nabla \cdot \vv_{n+1} - \nabla \cdot \vv_n) p_{n+1} \, d^3\vx =0
\end{equation}
and hence it is orthogonal to the divergence-free subspace. This notion of
orthogonality ensures that the projection step removes the component of
non-zero divergence in $\vv_*$ without introducing any additional contribution
to the solution in the space that is orthogonal to the
projection~\cite{chorin_fm}, which over time could create a spurious drift in
the solution.

\subsection{A projection method for quasi-static elastoplasticity}
Following the previous two sections, we conclude that there is close
correspondence between the elastoplastic system and the Navier--Stokes
equations for fluid flow. There is a correspondence between the variables
$(\bsig,\vv)$ in the elastoplastic system and the variables $(\vv,p)$ for fluid
flow. The limiting procedures that are employed, where the equations are scaled
to examine long times, are identical.

It is therefore natural to consider whether the projection method for the
incompressible Navier--Stokes equations can be adapted for simulating
quasi-static elastoplasticity. Suppose that $\bsig_n$ is a discretized stress
field after $n$ timesteps, and consider making a timestep of size $\Dt$. To
begin, an intermediate stress $\bsig_*$ is calculated by neglecting the total
rate-of-deformation term $\tC : \tD$ in Eq.~\ref{eq:constit}, so that
\begin{equation}
  \frac{\bsig_* - \bsig_n}{\Dt} = -\bsig_n \cdot \bome_n + \bome_n \cdot \bsig_n -(\vv_n \cdot \nabla) \bsig_n - \tC : \Dpl_n.
  \label{eq:sproj1}
\end{equation}
Assuming the velocity $\vv_{n+1}$ at the $(n+1)$th step can be calculated, and
consequently that the total deformation $\tD_{n+1}$ is known, then the stress
at the $(n+1)$th timestep is given by
\begin{equation}
  \frac{\bsig_{n+1} - \bsig_*}{\Dt} = \tC : \tD_{n+1}.
  \label{eq:sproj2}
\end{equation}
Taking the divergence of this equation and enforcing that $\nabla \cdot
\bsig_{n+1}=\vec{0}$ yields
\begin{equation}
  \nabla \cdot \bsig_* = - \Dt\, \nabla \cdot (\tC : \tD_{n+1}).
  \label{eq:sproj3}
\end{equation}
Eq.~\ref{eq:sproj3} is an algebraic system for the velocity $\vv_{n+1}$. It is
analogous to Eq.~\ref{eq:fproj3} for the fluid projection method, and will
involve second-order differential operators. It may also involve mixed
derivatives, and coupling between the components of velocity, but in principle
can be solved using standard numerical linear algebra techniques. As in the
fluid projection method, the boundary conditions for $\vv_{n+1}$ will be
problem-specific, but typical cases will have simple implementations: a
constant velocity boundary condition gives a Dirichlet condition on
$\vv_{n+1}$, while a traction boundary condition gives a Neumann-like condition
(discussed in Sec.~\ref{sec:free}). Once $\vv_{n+1}$ is calculated,
Eq.~\ref{eq:sproj2} can be used to evaluate $\bsig_{n+1}$.

A schematic representation of the algorithm is shown in Fig.~\ref{fig:schem}(b)
in the vector space $\vcsps$ of stresses, where the quasi-static solutions form
a subspace. As for the fluid projection method, it is useful to establish a
notion of orthogonality by introducing an inner product. This can be
constructed by making use of the compliance tensor $\tS$, which gives the
infinitesimal strain $\boldsymbol\epsilon$ in terms of stress according to
$\boldsymbol\epsilon=\ten{S}:\bsig$, so that $\tS=\tC^{-1}$. For real
materials, both $\tS$ and $\tC$ are positive-definite, in order to ensure that
the strain energy density is positive. For two stresses $\ten{a}, \ten{b} \in
\vcsps$, consider the inner product defined as
\begin{equation}
  \langle \ten{a}, \ten{b} \rangle = \int \ten{a} : \ten{S} : \ten{b} \, d^3\vx.
  \label{eq:sinprod}
\end{equation}
Since $\tS$ is positive-definite, this will be a valid inner product. The
projection $\bsig_\text{P}=\bsig_{n+1}-\bsig_*$ satisfies
\begin{eqnarray}
  \langle \bsig_{n+1} - \bsig_n, \Dsig \rangle &=& \Delta t \int (\bsig_{n+1} - \bsig_n ) : \ten{S} : (\tC : \tD_{n+1}) \, d^3\vx \nonumber \\
  &=&  \Delta t \int (\bsig_{n+1} - \bsig_n) : \tD_{n+1} \, d^3 \vx = \Delta t \int (\bsig_{n+1} - \bsig_n) : \nabla \vv_{n+1} \, d^3 \vx \nonumber \\
  &=& - \Delta t \int (\nabla \cdot \bsig_{n+1} - \nabla \cdot \bsig_n) \cdot \vv_{n+1} \, d^3 \vx = 0,
\end{eqnarray}
and therefore the projection is orthogonal the subspace of quasi-static
solutions. For an isotropic linear elastic material with bulk modulus $K$ and
shear modulus $\mu$ the components of the stiffness tensor are
\begin{equation}
  C_{ijkl} = \lambda \delta_{ij}\delta_{kl} + \mu(\delta_{ik}\delta_{jl} + \delta_{il}\delta_{jk}),
\end{equation}
where \smash{$\lambda=K-\frac{2\mu}{3}$} is L\'ame's first parameter. The
components of the compliance tensor are
\begin{equation}
  S_{ijkl} = \frac{1}{6K\mu} \left[ -\lambda \delta_{ij} \delta_{kl} + \tfrac{3K}{2} (\delta_{ik}\delta_{jl} + \delta_{il}\delta_{jk})\right].
\end{equation}
For this case, the inner product can be written as
\begin{equation}
  \langle \ten{a},\ten{b} \rangle = \frac{1}{6K \mu} \int (3K \ten{a} : \ten{b} - \lambda (\tr \ten{a})(\tr \ten{b}) )\,d^3 \vec{x}.
\end{equation}
As described in \ref{app:unique}, an integral argument can also be used to show
that Eq.~\ref{eq:sproj3} has a unique solution for Dirichlet boundary
conditions.

\section{A numerical implementation}
\label{sec:numerics}
We now describe a specific finite-difference numerical implementation of the
algorithms presented in Sec.~\ref{sec:methods}. We make use of a rate-dependent
elastoplastic model of a BMG that is based upon the STZ theory. Using this
model, we test the quasi-static time-integration method against the traditional
explicit scheme. All of the methods described below were implemented in a
custom-written C++ code, using the OpenMP library to multithread the loops
involved in the finite-difference update.

\subsection{Kinematics and elasticity}
A plane strain formulation in the $x$ and $y$ coordinates is
used~\cite{slaughter02}. The velocity is given by $\vv=(u,v,0)$, and the stress
tensor is written as
\begin{equation}
\bsig = \left(
\begin{array}{ccc}
  -p + s -q & \tau & 0 \\
  \tau & -p-s-q & 0 \\
  0 & 0 & -p + 2q \\
\end{array}
\right).
\end{equation}
Here, $p$ is the pressure, $s$ and $\tau$ are the components of deviatoric stress
within the $xy$ plane, and $q$ is the component of deviatoric stress in the $z$
direction out of the plane. The deviatoric part of the stress tensor is written
as \smash{$\bsig_0 = \bsig - \frac{1}{3}\ten{1} \tr \bsig$} and the magnitude
of the deviatoric stress tensor is
\smash{$|\bsig_0|=\bs=\sqrt{s^2+\tau^2+3q^2}$}. The density is assumed to be a
constant $\rho_0$, since elastic deformations are assumed to be small, and the
plastic deformation model is purely deviatoric. In component form,
Eq.~\ref{eq:force} reads
\begin{eqnarray}
  \label{eq:sys_start} \rho_0 \drt{u}&=&-\prx{p}-\prx{q}+\prx{s}+\pry{\tau}+\visc \nabla^2 u, \\
  \label{eq:vel_end} \rho_0 \drt{v}&=&-\pry{p}-\pry{q}-\pry{s}+\prx{\tau} + \visc \nabla^2 v,
\end{eqnarray}
where a small additional viscous stress term, $\visc \nabla^2 \vv$ has been
incorporated. This term is needed for numerical stability in the explicit
simulation method. However, it is not needed for numerical stability in the
quasi-static method.

The plastic deformation tensor is proportional to the deviatoric stress tensor
and can therefore be written as \smash{$\Dpl = \frac{\bsig_0}{\bar{s}}\dpl $},
where $\dpl$ is a scalar function described in detail in the following section.
In component form the constitutive equation, Eq.~\ref{eq:constit}, is given by
\begin{eqnarray}
  \drt{p} &=& -K \left(\prx{u} +\pry{v}\right), \\
  \drt{q} &=& -\frac{\mu}{3} \left(\prx{u} + \pry{v}\right) -\frac{2\mu q \dpl}{\bar{s}},\\
  \drt{s} &=& - 2\omega \tau + \mu\left( \prx{u} - \pry{v} \right) - \frac{2\mu s \dpl}{\bar{s}}, \\
  \label{eq:sys_end} \drt{\tau} &=& 2\omega s + \mu \left( \pry{u} + \prx{v} \right) - \frac{2\mu \tau \dpl}{\bar{s}},
\end{eqnarray}
where $\omega= (\p v /\p x - \p u / \p y)/2$, $K$ is the bulk modulus, and
$\mu$ is the shear modulus. Table~\ref{tab:param} shows the values of the
elastic parameters used in this study, which are based on Vitreloy 1, a
specific type of BMG whose mechanical properties have been well-studied.

\begin{table}
  \begin{center}
    \begin{tabular}{l|l}
      Parameter & Value \\
      \hline
      Young's modulus $E$ & 101~GPa\\
      Poisson ratio $\nu$ & 0.35 \\
      Bulk modulus $K$ & 122~GPa \\
      Shear modulus $\mu$ & 37.4~GPa\\
      Density $\rho_0$ &\smash{6125~kg~$\text{m}^{-3}$} \\
      Shear wave speed \smash{$c_s=\sqrt{\mu/\rho_0}$} & 2.47~km~$\text{s}^{-1}$
    \end{tabular}
  \end{center}
  \caption{Elasticity parameters used throughout the paper.\label{tab:param}}
\end{table}

\begin{table}
  \begin{center}
    \begin{tabular}{l|l}
      Parameter & Value \\
      \hline
      Yield stress $\sY$ & 0.85~GPa \\
      Molecular vibration timescale $\tau_0$ & $10^{-13}$~s\\
      Typical local strain $\epsilon_0$ & 0.3 \\
      Scaling parameter $c_0$ & 0.4 \\
      Typical activation barrier $\Delta/k_B$ & 8000~K\\
      Typical activation volume $\Omega$ & \smash{300~$\text{\AA}^3$} \\
      Bath temperature $T$ & 400~K \\
      Steady state effective temperature $\chi_\infty$ & 900~K \\
      STZ formation energy $e_z/k_B$ & 21000~K
    \end{tabular}
  \end{center}
  \caption{Parameter values for the STZ plasticity model used throughout the
  paper. The Boltzmann constant $k_B = 1.3806488 \times
  10^{-23}\text{~J~K}^{-1}$ is used to express the quantities $\Delta$ and
  $e_Z$ in terms of temperature.\label{tab:pparam}}
\end{table}

\subsection{Plasticity}
\label{sub:plasticity}
Plastic deformation is modeled using the shear transformation zone theory
of amorphous plasticity~\cite{falk98,langer08}. We employ a version of the
model used to study fracture~\cite{rycroft12b}, which is based on
recent theoretical developments~\cite{bouchbinder09,bouchbinder09b}, although
simplified to retain only the crucial details. Here, we sketch the theoretical
principles behind the model and provide the relevant equations.

Consider a BMG at a temperature $T$ below the glass transition temperature. If
no stress is applied, then the constituent atoms will undergo thermal
vibrations, but will largely remain in the same overall packing configuration
with their neighbors; in terms of an energy landscape, they are trapped within
a potential well representing one mechanically stable configuration. If the BMG
is subjected to a shear stress, then discrete events will occur whereby some
atoms in a local region undergo an irreversible change in configuration---the
applied stress changes the energy landscape to lower the potential barrier of
the well, so that it becomes possible to jump to another well representing a
different mechanically stable configuration.

This physical picture can be used to derive a continuum plasticity model. One
imagines that the material has a population of shear transformation zones,
which represent localized regions that are susceptible to shear-driven
configurational changes. The density of STZs is described in terms of an
effective disorder temperature $\chi$. For $\bs < \sY$, where $\sY$ is the
yield stress of the material, the plastic deformation is zero. For $\bs \ge
\sY$, the plastic deformation is given by
\begin{equation}
  \dpl(\bsig_0,T,\chi) = \frac{\Lambda(\chi) \sC(\bs,T)}{\tau_0} \left( 1 -
  \frac{\sY}{\bs} \right),
  \label{eq:stzdpl}
\end{equation}
where $\tau_0$ is a molecular vibration timescale, $\sC(\bs,T)$ is the STZ
transition rate, and \smash{$\Lambda(\chi)=e^{-e_z/k_B \chi}$} is the density of STZs
in terms of effective temperature, where $e_z$ is the STZ formation energy and
$k_B$ is the Boltzmann constant. The function $\sC(\bs,T)$ is specified in
terms of the forward and backward STZ transition rates,
\begin{equation}
\sC(\bs,T) = \tfrac{1}{2}(\sR(\bs,T)+\sR(-\bs,T)),
\label{eq:stzc}
\end{equation}
which follow a linearly stress-biased thermal activation process
\begin{equation}
\sR(\pm \bs, T) = \exp\left( - \frac{\Delta \mp \Omega \epsilon_0 \bs}{k_B T} \right),
\label{eq:stzr}
\end{equation}
where $\Delta$ is a typical energy activation barrier, $\Omega$ is a typical
activation volume, and $\epsilon_0$ is a typical local strain at the
transition. Substituting Eq.~\ref{eq:stzr} into Eq.~\ref{eq:stzc} yields
\begin{equation}
\sC(\bs,T) = e^{-\Delta/k_BT} \cosh \frac{\Omega\epsilon_0 \bs}{k_B T}.
\label{eq:stzceq}
\end{equation}
For very large positive values of $\bs$, it is possible that the stress-biasing
$\Omega \epsilon_0 \bs$ will exceed the activation barrier $\Delta$, in which
case the physical picture of a thermally activated process is no longer valid.
In previous work, we have assumed that for $\bs \Omega\epsilon_0 \ge \Delta$
the plastic behavior is dominated by a different, weaker, dissipative
mechanism~\cite{langer08,rycroft12}. However, we omit this term here for
mathematical simplicity. For the parameters given in Table~\ref{tab:pparam} the
barrier is reached at $\bs=1.44\sY$, and apart from the final example in
Subsec.~\ref{sub:eqtrans} where this issue is considered in more detail, the
deviatoric stresses never exceed $1.35\sY$, since the exponential growth of
$\dpl$ as a function of $\bs$ causes large deviatoric stresses to rapidly
relax. The effective temperature follows a heat equation of the form
\begin{equation}
  \label{eq:chiup}
  c_0\frac{d\chi}{dt} = \frac{(\Dpl : \bsig_0)(\chi_\infty - \chi) }{\sY}
\end{equation}
so that $\chi$ increases in response to plastic deformation and saturates at
$\chi_\infty$. Since an increase in $\chi$ will also increase $\dpl$ as given
by Eq.~\ref{eq:stzdpl}, the plasticity model typically leads to shear
banding~\cite{manning07,manning09}.

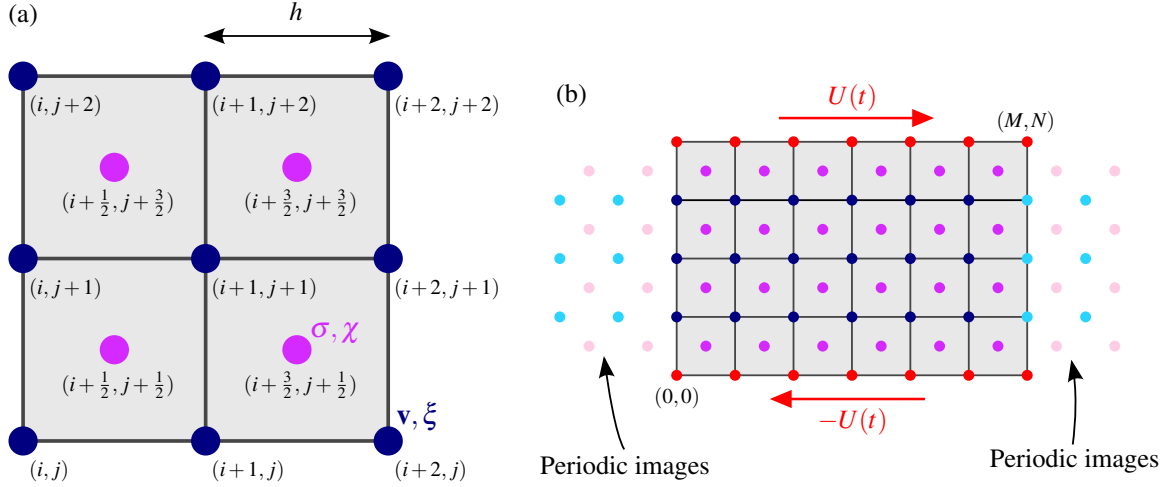
\begin{figure}
  \setlength{\unitlength}{0.86bp}
  \begin{center}
    {\footnotesize
    \begin{picture}(500,215)(0,0)
      \put(0,20){\includegraphics[scale=0.86]{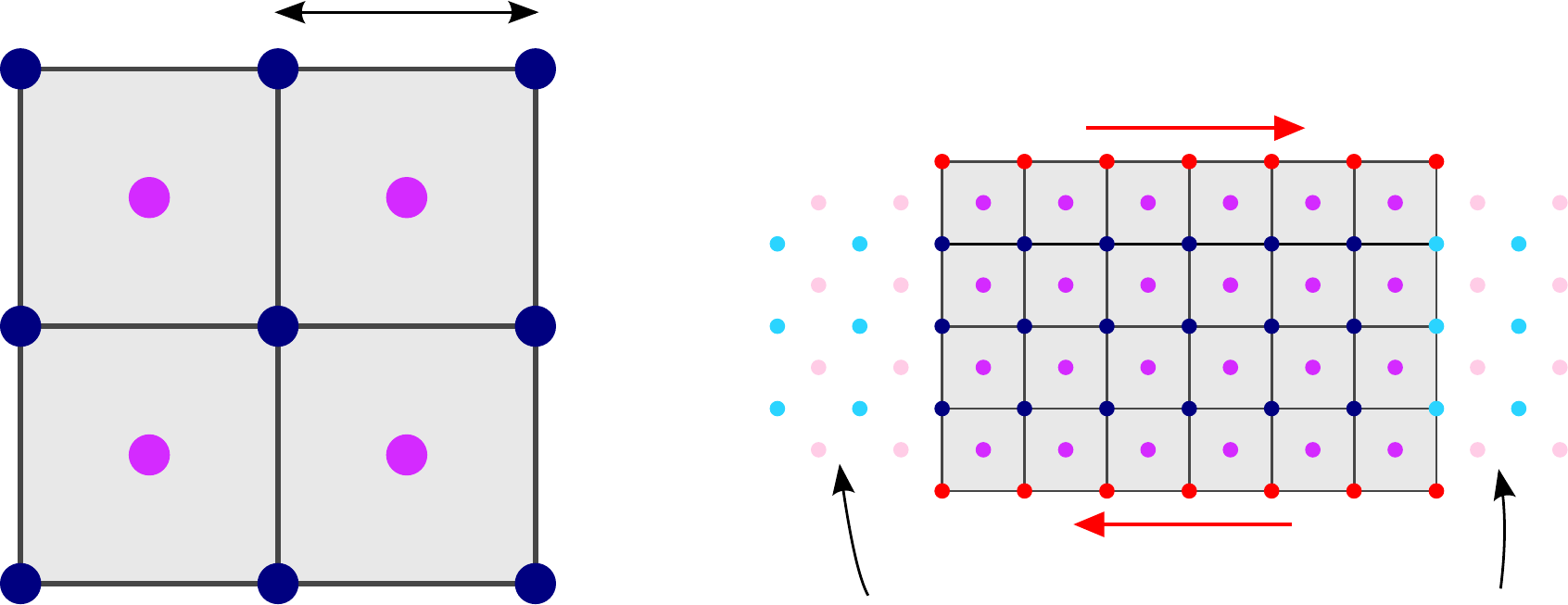}}
      \put(126,215){\makebox(0,0)[c]{$h$}}
      \put(170,34){\small \textcolor{dblue}{$\vv,\bxi$}}
      \put(132,72){\small \textcolor{dmag}{$\bsig,\chi$}}
      \put(370,177){\makebox(0,0)[c]{\textcolor{dred}{$U(t)$}}}
      \put(370,35){\makebox(0,0)[c]{\textcolor{dred}{$-U(t)$}}}
      \put(233,12){Periodic images}
      \put(430,15){Periodic images}
      \put(0,210){(a)}
      \put(240,175){(b)}
      \scriptsize
      \put(9,11){$(i,j)$}
      \put(89,11){$(i+1,j)$}
      \put(48,51){\makebox(0,0)[c]{$(i+\hf,j+\hf)$}}
      \put(128,51){\makebox(0,0)[c]{$(i+\thf,j+\hf)$}}
      \put(48,131){\makebox(0,0)[c]{$(i+\hf,j+\thf)$}}
      \put(128,131){\makebox(0,0)[c]{$(i+\thf,j+\thf)$}}
      \put(169,11){$(i+2,j)$}
      \put(9,91){$(i,j+1)$}
      \put(89,91){$(i+1,j+1)$}
      \put(169,91){$(i+2,j+1)$}
      \put(9,171){$(i,j+2)$}
      \put(89,171){$(i+1,j+2)$}
      \put(169,171){$(i+2,j+2)$}
      \put(283,43){$(0,0)$}
      \put(433,164){$(M,N)$}
    \end{picture}}
  \end{center}
  \vspace{3mm}
  \caption{(a) Arrangement of fields in the spatial discretization. The
  simulation is divided into square cells of side length $h$. The velocity
  $\vv$ and reference map $\bxi$ are stored at cell corners (dark blue), which
  are indexed with integers. The stress tensor $\bsig$ and effective
  temperature $\chi$ are stored at cell centers (magenta), which are indexed
  with half-integers. (b) Grid arrangement in the shearing simulation. The
  velocity in the top and bottom rows (red) of the simulation is fixed to
  create simple shear. To enforce the periodic boundary conditions in the
  horizontal $x$ direction, periodic images for both the cell-centered (pink)
  and cell-cornered (light blue) fields are used. In the example shown,
  $(M,N)=(6,4)$.\label{fig:grid} }
\end{figure}

\subsection{Numerical methods for explicit simulations}
\label{sub:directsim}
The simulations are carried out on a rectangular $M \times N$ grid of square
cells with side length $h$. As shown in Fig.~\ref{fig:grid}(a), a staggered
arrangement is used whereby the components of velocity $u$, $v$ are stored at
cell corners and indexed with integers, and the components of stress $p$, $q$,
$s$, $\tau$ and effective temperature $\chi$ are stored at cell centers and
indexed with half-integers. The explicit simulation method employs
Eqs.~\ref{eq:sys_start} to \ref{eq:sys_end} and Eq.~\ref{eq:chiup} to
explicitly update all the simulation fields, using a first-order temporal
discretization and a second-order spatial discretization.

The first derivatives on the right hand sides of Eqs.~\ref{eq:sys_start} to
\ref{eq:sys_end} are evaluated using centered differencing. It can be observed
that the equations for velocity depend on first derivatives of stress and vice
versa. If $f_{i,j}$ represents one of the discretized fields at a given
instant, then the staggered first derivative in the $x$ direction is evaluated
as
\begin{equation}
  \label{eq:centdiff}
  \left[\frac{\p f}{\p x}\right]_{i+\hf,j+\hf} = \frac{f_{i+1,j}+f_{i+1,j+1} - f_{i,j}-f_{i,j+1}}{2h}.
\end{equation}
The viscosity terms make use of a colocated second-order derivative, which is
evaluated in the $x$ direction as
\begin{equation}
  \label{eq:centdiff2}
  \left[\frac{\p^2 f}{\p x^2}\right]_{i,j} = \frac{f_{i+1,j} - 2f_{i,j} + f_{i-1,j}}{h^2}.
\end{equation}
The advective derivatives on the left hand side of Eqs.~\ref{eq:sys_start} to
\ref{eq:sys_end} need to be upwinded for stability. This is achieved by using
the second-order ENO numerical scheme~\cite{shu88}, which in the $x$ direction
is given by
\begin{equation}
  \label{eq:eno}
  \left\{\frac{\p f}{\p x}\right\}_{i,j} = \frac{1}{2h}\left\{
  \begin{array}{ll}
    -f_{i+2,j}+4f_{i+1,j}-3f_{i,j} & \qquad \text{if $u_{i,j}<0$ and $|[f_{xx}]_{i,j}| > |[f_{xx}]_{i+1,j}|$,} \\
      3f_{i,j}-4f_{i-1,j}+f_{i-2,j} & \qquad \text{if $u_{i,j}>0$ and $|[f_{xx}]_{i,j}| > |[f_{xx}]_{i-1,j}|$,} \\
    f_{i+1,j}-f_{i-1,j}& \qquad \textrm{otherwise,}
  \end{array}
  \right.
\end{equation}
where $[f_{xx}]_{i,j}$ is the second-order centered-difference at $i,j$
evaluated using Eq.~\ref{eq:centdiff2}. The ENO derivative therefore switches
between an upwinded one-sided derivative and a centered derivative, depending
on which set of three field values is more colinear. In the $y$ direction,
analogous expressions to Eq.~\ref{eq:centdiff} and Eq.~\ref{eq:eno} are used.

The first-order forward Euler scheme is used for timestepping. If velocity
components and pressure at timestep $n$ are written as $u_n$, $v_n$, and $p_n$,
and a timestep $\Delta t$ is taken, then at timestep $(n+1)$ they are given by
\begin{eqnarray}
  \label{eq:direct_start}
  \rho_0 \frac{u_{n+1}- u_n}{\Dt} &=& - \rho_0 \adv u_n -\prx{p_n}-\prx{q_n}+\prx{s_n}+\pry{\tau_n}+\visc \nabla^2 u_n, \\
  \label{eq:vel2}
  \rho_0 \frac{v_{n+1}-v_n}{\Dt} &=& - \rho_0 \adv v_n -\pry{p_n}-\pry{q_n}-\pry{s_n}+\prx{\tau_n} + \visc \nabla^2 v_n, \\
  \frac{p_{n+1} - p_n}{\Dt} &=& -\adv p_n - K \left(\prx{u_n} +\pry{v_n}\right).
\end{eqnarray}
The deviatoric stress components are updated with a similar procedure, but make
use of a modification to accommodate for the rapid growth of $\dpl$ when $\bs$
exceeds the yield stress $\sY$, which causes a loss of accuracy if $\Dt$ is too
large. Suppose that at a given location and timestep, a discretized deviatoric
stress $\bs_n$ is slightly above $\sY$. Physically, plastic deformation should
cause the deviatoric stress to decrease until reaching the yield surface so
that $\bs_{n+1}\approx\sY$. However, if other terms are neglected, then the
Euler step will give $\bs_{n+1} = \bs_n - 2\mu\dpl \Dt$ at the next
timestep, which could be substantially lower than $\sY$ if $\dpl$ is large,
overshooting the yield surface. To solve this, an adaptive timestepping routine
is used that divides the interval $\Dt$ into subintervals so that the
incremental changes to $\bs$ remain small---this accomplishes a similar goal as
the return-mapping algorithms for rate-independent
plasticity~\cite{wilkins63,simo98}. The routine, described in \ref{app:adapt},
considers the coupled system $\bs$ and $\chi$ and returns modified functions
$\tdpln$ and $\tilde{F}_n$ for use in the finite-difference update. The
deviatoric stress and effective temperature are updated according to
\begin{eqnarray}
  \label{eq:direct_sstart}
  \frac{q_{n+1}-q_n}{\Dt } &=& -\adv q_n - \frac{\mu}{3} \left(\prx{u_n} +\pry{v_n}\right) - \frac{2 \mu \tdpln q_n}{\bs_n}, \\
  \frac{s_{n+1}-s_n}{\Dt } &=& -\adv s_n - 2\omega_n \tau_n +  \mu \left( \prx{u_n} - \pry{v_n} \right) - \frac{2 \mu \tdpln s_n}{\bs_n}, \\
  \label{eq:direct_send}
  \frac{\tau_{n+1}-\tau_n}{\Dt } &=& -\adv \tau_n + 2\omega_n s_n + \mu \left( \pry{u_n} + \prx{v_n} \right)- \frac{2 \mu \tdpln \tau_n}{\bs_n}, \\
  \label{eq:direct_end} \frac{\chi_{n+1} - \chi_n}{\Dt} &=& - \adv \chi_n + \tilde{F}_n,
\end{eqnarray}
where $\omega_n= (\p v_n /\p x - \p u_n / \p y)/2$ and the $\vv_n$ term in the
advective derivatives is evaluated as the average of the velocities at the four
corners of the grid cell.

The simulation also makes use of a reference map vector field
$\bxi=(\xi^x,\xi^y)$ stored at cell corners. This field has no physical
influence, but is used to track the deformation of the material. It is
initialized as
\begin{equation}
  \label{eq:refmapinit}
  \bxi(\vx,0)=\vx
\end{equation}
and is then updated according to
\begin{equation}
  \frac{d\bxi}{dt} = \frac{\p \bxi}{\p t} + (\vv\cdot\nabla) \bxi = \vec{0},
\end{equation}
following the same discretization methods as for the other fields. Contours of
the components of the reference map initially form a rectangular grid and then
deform with the material. Using $\bxi$, the $(2\times 2)$-component deformation
gradient tensor is given by
\begin{equation}
  \label{eq:defgradient}
  \vec{F}= \frac{\p \vx}{\p \bxi},
\end{equation}
which can be numerically evaluated using centered differences of $\bxi$. Once
$\vec{F}$ is known, the Green--Saint-Venant strain tensor is given by
\smash{$\vec{E} = \frac{1}{2}(\vec{F}^\transpose\vec{F}-\vec{1})$}. The deviatoric
part of the strain tensor is defined as \smash{$\vec{E}_0= \vec{E} -
\frac{1}{2}\vec{1}\tr \vec{E}$}.

\subsection{Numerical methods for quasi-static simulations}
\label{sub:quasisim}
The quasi-static scheme makes use of the same simulation framework as the
explicit scheme. It employs the same rectangular grid, and uses
Eqs.~\ref{eq:centdiff} and \ref{eq:eno} for carrying out spatial derivatives.
To carry out a timestep of size $\Dt$, Eq.~\ref{eq:sproj1} is first used to
calculate an intermediate stress $\bsig_*$, which in component form is
\begin{eqnarray}
  \frac{p_*-p_n}{\Dt } &=& -\adv p_n, \\
  \label{eq:qs_sstart}
  \frac{q_*-q_n}{\Dt } &=& -\adv q_n - \frac{2 \mu \tdpln q_n}{\bs_n}, \\
  \frac{s_*-s_n}{\Dt } &=& -\adv s_n - 2\omega_n \tau_n - \frac{2 \mu \tdpln s_n}{\bs_n}, \\
  \label{eq:qs_send}
  \frac{\tau_*-\tau_n}{\Dt } &=& -\adv \tau_n + 2\omega_n s_n - \frac{2 \mu \tdpln \tau_n}{\bs_n}.
\end{eqnarray}
The adaptive procedure described in \ref{app:adapt} is used to evaluate the
plastic deformation term \smash{$\tdpln$} that features in these equations. It
also returns $\tilde{F}_n$, which allows $\chi_{n+1}$ to be calculated
according to Eq.~\ref{eq:direct_end}.

If the velocity $\vv_{n+1}$ at timestep $n+1$ is known, then by following
Eq.~\ref{eq:sproj2}, the components of $\bsig_{n+1}$ are given by
\begin{eqnarray}
  \label{eq:proj_start} \frac{p_{n+1}-p_*}{\Dt} &=& - K \left(\prx{u_{n+1}} +\pry{v_{n+1}}\right), \\
  \frac{q_{n+1}-q_*}{\Dt} &=& - \frac{\mu}{3} \left(\prx{u_{n+1}} +\pry{v_{n+1}}\right), \\
  \frac{s_{n+1}-s_*}{\Dt} &=& \mu \left( \prx{u_{n+1}} - \pry{v_{n+1}} \right),  \\
  \label{eq:proj_end} \frac{\tau_{n+1}-\tau_*}{\Dt} &=& \mu \left( \pry{u_{n+1}} + \prx{v_{n+1}} \right).
\end{eqnarray}
To calculate $\vv_{n+1}$, the quasi-staticity constraint at the $(n+1)$th
timestep is used, which by retaining the viscous stress is slightly modified
to $\vec{0} = \nabla\cdot \bsig_{n+1} + \visc\nabla^2 \vv_{n+1}$. Following
Eq.~\ref{eq:sproj3}, the velocity satisfies
{\small
\begin{eqnarray}
  \label{eq:double_multi1} (\mu+K'+\visc') \frac{\p^2 u_{n+1}}{\p x^2} + (\mu+\visc') \frac{\p^2 u_{n+1}}{\p y^2} + K' \frac{\p^2 v_{n+1}}{\p x \p y} &=& \frac{1}{\Dt } \left(\prx{p_*}+\prx{q_*}-\prx{s_*}-\pry{\tau_*}\right), \\
  \label{eq:double_multi2} (\mu+\visc') \frac{\p^2 v_{n+1}}{\p x^2} + (\mu+K'+\visc') \frac{\p^2 v_{n+1}}{\p y^2} + K' \frac{\p^2 u_{n+1}}{\p x \p y}&=& \frac{1}{\Dt } \left( \pry{p_*}+\pry{q_*}+\pry{s_*}-\prx{\tau_*} \right),
\end{eqnarray}}where \smash{$K'=K+\frac{\mu}{3}$} and \smash{$\visc'=\frac{\visc}{\Dt}$}. In
the typical regime of interest where $\Dt$ becomes large, the effect of the
viscous term is therefore negligible.

Eqs.~\ref{eq:double_multi1} and \ref{eq:double_multi2} form an algebraic system
for the components of velocity. The system features second derivatives and
bears some similarity to the Poisson equation that must be solved for the fluid
projection method. However, the system is more complicated, since the two
components of velocity are coupled, and a mixed $xy$-derivative is present. To
solve the equations, a linear system $A_0$ is constructed where the derivatives
are discretized using Eqs.~\ref{eq:centdiff} \& \ref{eq:centdiff2}, and
\begin{equation*}
  \left[\frac{\p^2f}{\p x \p y}\right]_{i,j} = \frac{f_{i+1,j+1} - f_{i+1,j-1} - f_{i-1,j+1} + f_{i-1,j-1}}{4h^2},
\end{equation*}
where $f_{i,j}$ represents the components of an arbitrary field. The linear
system also takes into account problem-specific boundary conditions, which are
discussed later.

The presence of the mixed derivative means that the linear system is not weakly
diagonally dominant, unlike the Poisson problem for the fluid projection
method. However, in general as discussed previously, the matrix will be
symmetric and positive-definite, other than possible complications due to the
application of boundary conditions. The linear system is therefore well-suited
to be solved by many linear algebra techniques and will admit a unique
solution. For the cases considered here, the linear system is solved using a
custom-written geometric multigrid algorithm.

\section{Shearing between two parallel plates}
\label{sec:shearing}
The first example considered is a material being sheared between two parallel
plates. This example has simple boundary conditions, but exhibits complex
behavior and shear banding, making it a useful environment in which to compare
the explicit and quasi-static simulation approaches. The example uses a domain
that is periodic in the $x$ direction and covers $-\gamma L < x \le \gamma L,
-L\le y \le L$ where $\gamma$ is a dimensionless constant. Initially, the
velocity and Cauchy stress are zero, and the reference map is given by
Eq.~\ref{eq:refmapinit}. A natural time unit is $\tsca=L/c_s$. The boundary
conditions on the top and bottom boundaries are
\begin{equation}
  \label{eq:shearbc}
  \vv(x,\pm L,t) = (\pm U(t),0), \qquad \left.\frac{\p \bsig}{\p y}\right|_{y=\pm L} =\left.\frac{\p \chi}{\p y}\right|_{y=\pm L} =0, \qquad \bxi(x,\pm L,t) = (x\mp X(t),\pm L),
\end{equation}
where the function $U(t)$ satisfies
\begin{equation}
U(t) = \left\{
\begin{array}{ll}
  \frac{U_\text{B}t}{\tsca} & \qquad \text{for $t<\tsca$,} \\
  U_\text{B} & \qquad \text{for $t\ge \tsca$,}
\end{array}
\right.
\label{eq:uform}
\end{equation}
so that the speed of the parallel plates is linearly increased to a value
$U_\text{B}$, after which it remains constant. This form for $U(t)$ causes the
stresses in the material to gradually increase, and avoids the problem that
applying $U(t)=U_\text{B}$ for $t>0$ would immediately create a very large
deformation rate next to the boundaries. For consistency, the function $X(t)$
in Eq.~\ref{eq:shearbc} is given by
\begin{equation}
X(t) = \int_0^t U(t') dt' = \left\{
\begin{array}{ll}
  \frac{U_\text{B}t^2}{2\tsca} & \qquad \text{for $t<\tsca$,} \\
  U_\text{B}\left(t-\frac{\tsca}{2}\right) & \qquad \text{for $t\ge \tsca$.}
\end{array}
\right.
\end{equation}
A schematic of the grid point layout is shown in Fig.~\ref{fig:grid}(b). The
cell-cornered grid points $(i,j)$ cover the index ranges $i=0, 1, \ldots, M-1$
and $j=0, 1, \ldots, N$, and cell-centered grid points cover the index ranges
\smash{$i=\frac{1}{2},\frac{3}{2},\ldots,\frac{2M-1}{2}$} and
\smash{$j=\frac{1}{2},\frac{3}{2},\ldots,\frac{2N-1}{2}$}. The location of grid
point $(i,j)$ is at \smash{$(x,y)=(-\gamma L+hi,-L+hj)$} so that $j=0$ is
located on the bottom boundary and $j=N$ is located on the top boundary.
Throughout the simulation, the field values for $j=0$ and $j=N$ are set using
the boundary conditions in Eq.~\ref{eq:shearbc}.

Explicit and quasi-static simulations are carried out using the methods
described in Subsecs.~\ref{sub:directsim} and \ref{sub:quasisim} respectively, and
are applied to grid points in the range \smash{$\frac{1}{2} \le j \le
\frac{2N-1}{2}$}. To handle the periodic boundary conditions, the spatial
finite-difference operators wrap around; for example, a reference to an
arbitrary field value $f_{M,j}$ is treated as $f_{0,j}$. In addition, a
displacement of $2\gamma L$ is applied to the $x$-component of the reference
map, so that $\xi^x_{M,j} = \xi^x_{0,j}+2\gamma L$. When calculating upwinded
derivatives in the $y$-direction at \smash{$j=\frac{1}{2},1$} and
\smash{$j=n-1,\frac{2n-1}{2}$} using Eq.~\ref{eq:eno}, the simulation falls
back on a first-order upwinded derivative if not enough grid points are
available to calculate the ENO discretization. For this example, the algebraic
problem considered in the quasi-static simulation method is simple to implement
and makes use of Dirichlet conditions on $\vec{v}$ at $j=0$ and $j=N$.

\subsection{Comparison of explicit and quasi-static methods}
We first consider a case where the parameters are chosen to allow for a
quantitative comparison between the explicit and quasi-static simulation
approaches. We make use of $L=1\text{~cm}$, $\gamma=4$, and consider an initial
effective temperature distribution of the form
\begin{equation}
  \chi(\vx,t) = 630\text{~K} + (170\text{~K})\exp\left(-\frac{|20\vx|^2}{2L^2}\right),
\end{equation}
corresponding to a small imperfection in the center of the domain. When
subjected to shear, we expect that a shear band will nucleate from the
imperfection, creating a region where plastic deformation will be localized.
The parameters given in Tables~\ref{tab:param} and \ref{tab:pparam} are used as
a baseline, and for the given value of $L$, the natural timescale is $\tsca=
4.05 \text{~\textmu{}s}$. A grid size of $640\times 160$ is used, so that the grid
spacing is \smash{$h=\frac{L}{80}$}.

To quantitatively compare the explicit and quasi-static simulation approaches,
a parameter $\zeta$ is introduced that can control the overall speed of the
dynamics in a manner similar to the scaling argument in
Eq.~\ref{eq:long_time_scale}. The boundary speed is set to $U_\text{B} =
10^{-7} \zeta L/\tsca = 247\zeta~\text{\textmu{}m/s}$ and the plastic
deformation rate is scaled by $\zeta$, by replacing $\tau_0$ with
$10^{-13}\zeta^{-1}\text{~s}$. Simulations over a duration of $2 \times
10^6 \tsca\zeta^{-1} = 8.09\zeta^{-1}\text{~s}$ are carried out, after which
the boundaries are each displaced by approximately $2\text{~mm}$. For
$\zeta=1$, the scales are approximately in physically reasonable ranges for
typical experimental tests. The viscous stress constant is
$\kappa=0.02L^2/\tsca$. The timestep used in the explicit simulation is
\smash{$\Delta t= \frac{\tsca h^2}{2L^2}$} so that the viscous stress can be
properly resolved. The timestep used in the quasi-static simulation is $\Delta
t = \frac{100\tsca}{\zeta}$.

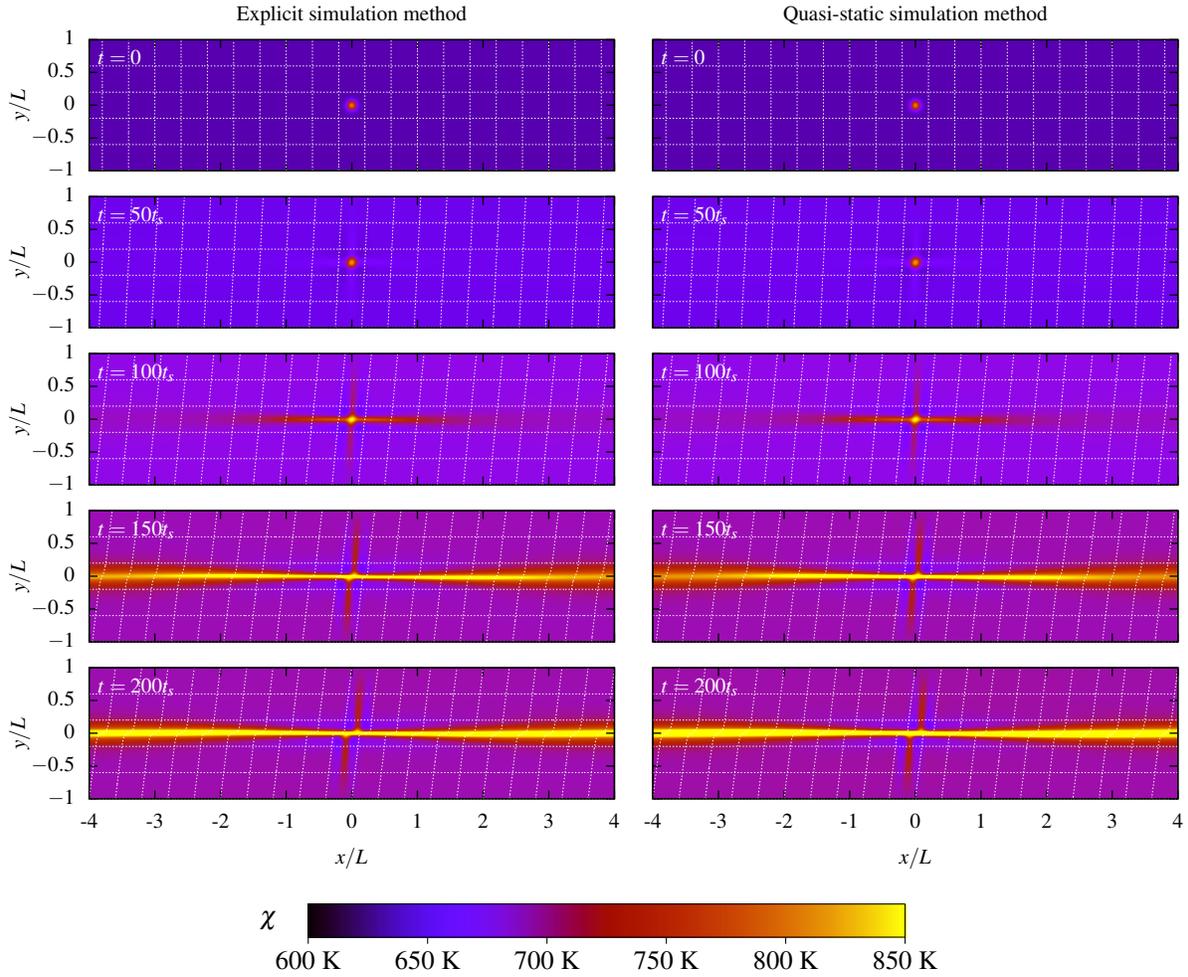
\begin{figure}
  \begin{center}
    {\scriptsize
    \include{basic_tem}}
    \vspace{-1mm}
    \setlength{\unitlength}{0.0125bp}
    \begin{picture}(21100,2000)(0,0)
      \footnotesize
      \put(1600,2000){\includegraphics[width=226pt,height=12.5pt]{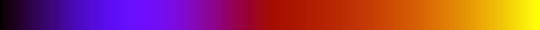}}
      \put(1600,2000){\line(1,0){18000}}
      \put(1600,1800){\line(0,1){1200}}
      \put(1600,3000){\line(1,0){18000}}
      \put(19600,1800){\line(0,1){1200}}
      \put(400,2500){\makebox(0,0)[c]{$\chi$}}
      \put(5200,1800){\line(0,1){200}}
      \put(8800,1800){\line(0,1){200}}
      \put(12400,1800){\line(0,1){200}}
      \put(16000,1800){\line(0,1){200}}
      \put(1600,1300){\makebox(0,0)[c]{600~K}}
      \put(5200,1300){\makebox(0,0)[c]{650~K}}
      \put(8800,1300){\makebox(0,0)[c]{700~K}}
      \put(12400,1300){\makebox(0,0)[c]{750~K}}
      \put(16000,1300){\makebox(0,0)[c]{800~K}}
      \put(19600,1300){\makebox(0,0)[c]{850~K}}
    \end{picture}
  \end{center}
  \caption{Plots of effective temperature $\chi$ at five time points for the
  shear band nucleation simulation, using the explicit simulation method (left)
  and the quasi-static simulation method (right). The thin dashed white lines
  are the contours of the components of the reference map $\bxi$, and show how
  the material deforms. As described in the text, the simulation is speeded up
  by a factor of $\zeta=10^4$ from physical parameters to make it
  computationally feasible to compare the two numerical
  methods.\label{fig:basic_tem}}
\end{figure}

Figure~\ref{fig:basic_tem} shows a sequence of snapshots of effective
temperature, for both the explicit simulation and the quasi-static simulation,
using an artificial scaling factor of $\zeta=10^4$. The two simulation methods
give very similar results and are hard to differentiate by eye. At
$t=50\tsca$, the effective temperature has increased uniformly by a small
amount throughout the material, but bands of slightly higher $\chi$ have begun
to emerge in the orthogonal directions from the initial imperfection. By
$t=100\tsca$, the horizontal band starts to dominate, and by $t=150\tsca$ it
has grown across the entire width of the simulation. The shear band continues
to grow larger by $t=200\tsca$, and accommodates most of the plastic
deformation.

\begin{figure}
  \begin{center}
    {\scriptsize
    \include{basic_p}}
    \setlength{\unitlength}{0.0125bp}
    \vspace{-1mm}
    \begin{picture}(21100,2000)(0,0)
      \footnotesize
      \put(1600,2000){\includegraphics[width=226pt,height=12.5pt]{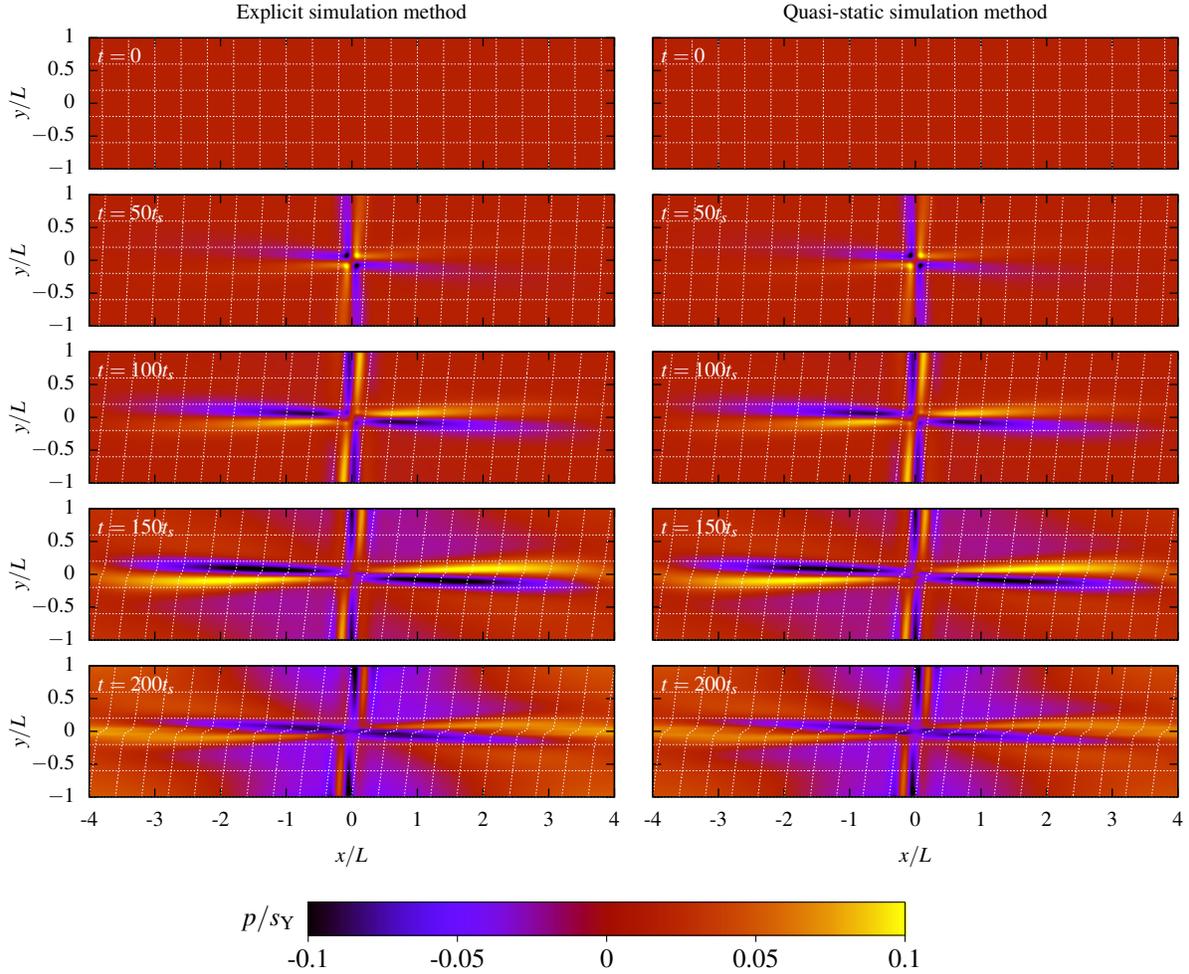}}
      \put(1600,2000){\line(1,0){18000}}
      \put(1600,1800){\line(0,1){1200}}
      \put(1600,3000){\line(1,0){18000}}
      \put(19600,1800){\line(0,1){1200}}
      \put(400,2500){\makebox(0,0)[c]{$p/\sY$}}
      \put(6100,1800){\line(0,1){200}}
      \put(10600,1800){\line(0,1){200}}
      \put(15100,1800){\line(0,1){200}}
      \put(1600,1300){\makebox(0,0)[c]{-0.1}}
      \put(6100,1300){\makebox(0,0)[c]{-0.05}}
      \put(10600,1300){\makebox(0,0)[c]{0}}
      \put(15100,1300){\makebox(0,0)[c]{0.05}}
      \put(19600,1300){\makebox(0,0)[c]{0.1}}
    \end{picture}
  \end{center}
  \caption{Plots of pressure $p$ at five time points for the shear band
  nucleation simulation, using the explicit simulation method (left) and the
  quasi-static simulation method (right). The thin dashed white lines are the
  contours of the components of the reference map $\bxi$, and show how the
  material is deforms. The simulation is speeded up by a factor of $\zeta=10^4$
  from physical parameters.\label{fig:basic_p}}
\end{figure}

When the full shear band initially forms at $t\approx 150\tsca$, it is
approximately three simulation grid points across, and may therefore not be
fully resolved; its width may partly be governed by numerical diffusion.
At later times as more plastic deformation occurs, the shear band width
continues to grow, consistent with one-dimensional studies~\cite{manning07}.
Figure~\ref{fig:basic_p} shows plots of the pressure field for the two
simulation methods, at the same sequence of time points. The pressure fields
are relatively small, reaching values up to \smash{$\frac{1}{10}\sY$}, but
again there is very good agreement between the two methods. The increased
plastic deformation near the initial imperfection leads to a small quadrupolar
feature the pressure field.

\begin{figure}
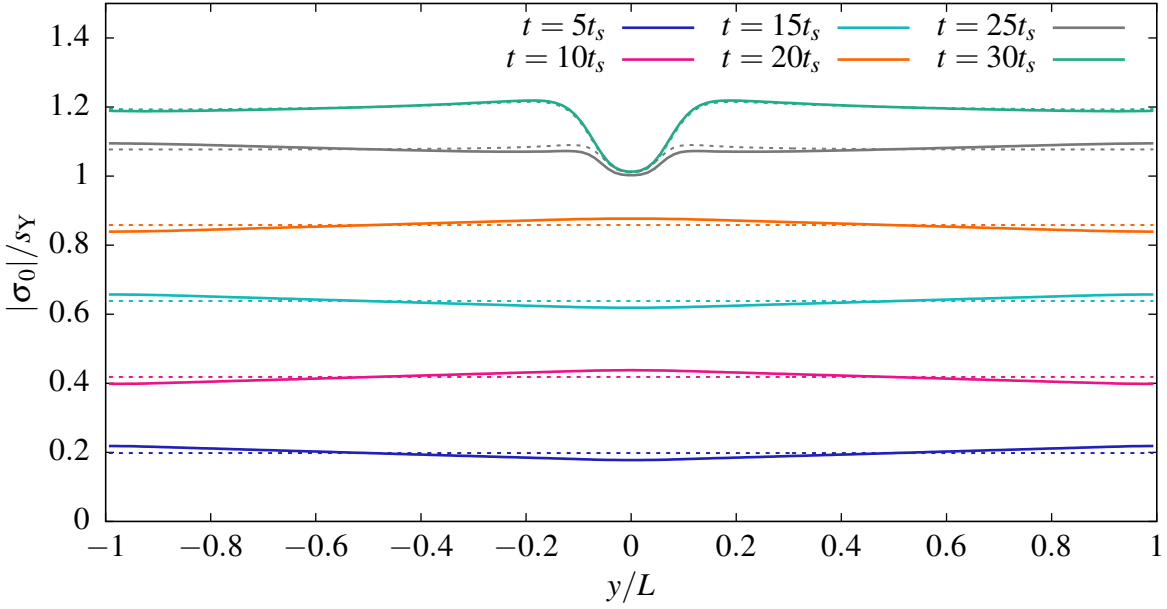

  \include{dev_tran_comp}
  \caption{Comparison of the cross-sections of the deviatoric stress field on
  the line $x=0$ for the explicit shear band simulation (solid lines) and the
  quasi-static shear band simulation (dashed lines).~\label{fig:dev_tran_comp}}
\end{figure}

Figure~\ref{fig:dev_tran_comp} shows the cross sections of the deviatoric
stress $|\bsig_0|$ for the two simulations, for several time points up to
$t=30\tsca$. The graph highlights some small differences between the methods.
In the quasi-static simulation, $|\bsig_0|$ is uniform in $y$ up to $t=20\tsca$
while the material is in the elastic regime and the shear stress is below the
yield stress. The corresponding plots for the explicit simulation are similar,
although show slight oscillations, due to elastic waves propagating across the
material. Even though the shearing velocity is gradually increased following
Eq.~\ref{eq:uform}, some small elastic waves are introduced at the start of the
simulation, which continue to propagate across the simulation since there is
little damping to remove them. By $t=25\tsca$ some plastic deformation starts
to occur resulting in a reduction of shear stress near $y=0$. Since the plastic
deformation introduces some dissipation, the elastic waves in the explicit
simulation are damped out, meaning that by $t=30\tsca$ the two simulation
methods come into closer agreement.

These simulations were carried out using eight threads on a Mac Pro (Late 2013)
with an 8-core 3~GHz Intel Xeon E5 processor. The explicit simulation used
2,560,000 timesteps and took a total wall clock time of $7578\text{~s}$,
corresponding to an average wall clock time of $2.96~\text{ms}$ per integration
step. The quasi-static simulation used 20,000 timesteps and took a total wall
clock time of $1378\text{~s}$, corresponding to an average wall clock time of
$68.9~\text{ms}$ per integration step. While the quasi-static simulation step
takes more than twenty times longer than the explicit timestep due to solving a
linear system using the multigrid method, its ability to take much larger steps
means that the total simulation time is less than a fifth of the time for the
explicit simulation. At lower values of $\zeta$, the quasi-static simulation
will require the same computation time, while the computation time for the
explicit simulation take longer, since the time required is inversely
proportional to $\zeta$.

\subsection{Quantitative comparison of the explicit and quasi-static simulation methods}
The quasi-static system of equations given by Eqs.~\ref{eq:constit} and
\ref{eq:quasis} emerges from taking a limit of slow velocity and long times,
and in this section we quantitatively compare the two simulation methods in
this limit. We employ the same boundary conditions as in the previous section,
and we expect that as $\zeta$ is reduced, the differences between the two
methods will tend to zero. However, quantitatively examining this poses some
difficulties, since in addition to simulating different equations, the two
methods introduce different discretization errors. It is therefore necessary to
consider additional parameters that affect the discretization.

\begin{figure}
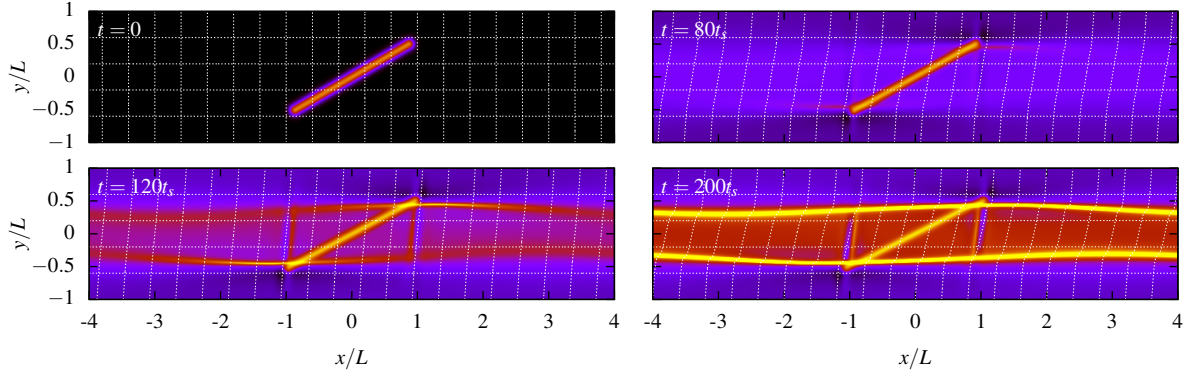

  \begin{center}
    {\scriptsize \include{qline}}
  \end{center}
  \caption{Four snapshots of the effective temperature $\chi$ field for a
  quasi-static simulation where a line of higher $\chi$ is initially introduced
  at an angle of $30^\circ$ relative to the horizontal. The simulation
  parameters are speeded up by a factor of $\zeta=10^4$ from physically
  realistic values. The color gradient is the same as that used in
  Fig.~\ref{fig:basic_tem}.\label{fig:qline}}
\end{figure}

To evaluate the differences between the explicit and quasi-static simulations,
a norm
\begin{equation}
  || \vec{f} || = \sqrt{ \frac{1}{16L^2} \int_{-4L}^{4L} dx \int_{-L}^{L} |\vec{f}|^2 dy}
  \label{eq:norm}
\end{equation}
is introduced where $\vec{f}$ is an arbitrary field, and the integrals are
evaluated using the trapezoidal rule. By interpreting $|\vec{f}|^2$
appropriately, Eq.~\ref{eq:norm} can be applied to scalars, vectors, and
tensors. To create more of a spread in the effective temperature field, we
consider an alternative initial condition describing a rotated line of
higher $\chi$. The function
\begin{equation}
\Gamma(x',y') = \left\{
\begin{array}{ll}
  \exp\left(-\frac{|20y'|^2}{2L^2}\right) & \qquad \text{if $|x'| \le L$,} \\
  \exp\left(-\frac{400( (|x'|-L)^2+y'^2)}{2L^2}\right) & \qquad \text{if $|x'| >L$,}
\end{array}
\right.
\end{equation}
is first introduced, after which the initial effective temperature is given by
\begin{equation}
  \chi(\vx,t) = \chi_0 + (800\text{~K}-\chi_0)\Gamma'(x\cos 30^\circ + y\sin 30^\circ, -x\sin 30^\circ + y\cos 30^\circ),
  \label{eq:line_chi0}
\end{equation}
where $\chi_0=600\text{~K}$. The direct timestep is \smash{$\Delta t=
\frac{\tsca h^2}{2L^2}$} as in the previous section, and a quasi-static
timestep of $\Delta t = \frac{200\tsca}{\zeta}$ is used as a baseline.
Figure~\ref{fig:qline} shows several snapshots of the effective temperature
field using the quasi-static method, where the boundary conditions are set
using $\zeta=10^4$. Shear bands nucleate from the ends of the line and grow
horizontally, although they follow slightly curved paths. By $t=200\tsca$ the
region between the two shear bands has undergone a substantial increase in
$\chi$.

\begin{figure}
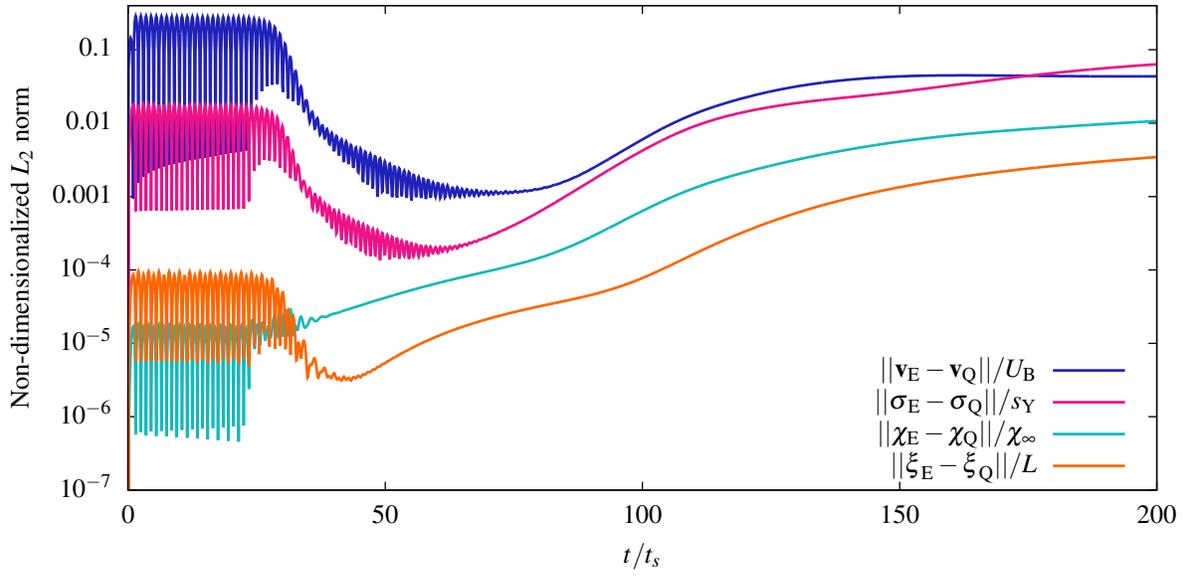

  \begin{center}
    {\footnotesize \include{l2norm}}
  \end{center}
  \caption{Non-dimensionalized differences between the simulation fields in the
  quasi-static and explicit simulations of the rotated line configuration shown
  in Fig.~\ref{fig:qline}, quantified using the $L_2$ norm defined in the
  text.\label{fig:l2diff}}
\end{figure}

A corresponding explicit simulation was carried out and four
non-dimensionalized norms $||\vv_\ted-\vv_\teq||/U_\text{B}$,
$||\bsig_\ted-\bsig_\teq||/\sY$, $||\chi_\ted-\chi_\teq||/\chi_\infty$, and
$||\bxi_\ted-\bxi_\teq||/L$ were evaluated at intervals of $0.2\tsca$, where the
subscripts of E and Q refer to the explicit and quasi-static simulation fields
respectively. The norms provide a measure of the global differences between the
fields, and the normalizing factors are chosen to make the fields in each norm
approximately of order unity. Plots of the differences in these fields are
shown in Fig.~\ref{fig:l2diff}. Throughout the simulation, all fields remain in
good agreement. The largest discrepancies are in the initial interval from $0
\le t < 25\tsca$, where all four norms exhibit oscillations. This is due to
elastic waves propagating across the explicit simulation, as discussed for
Fig.~\ref{fig:dev_tran_comp}. Once plastic deformation starts to occur at
$t\approx 25\tsca$ these oscillations are damped out, and the agreement between
stresses and velocities is improved by two orders of magnitude. Beyond
$t=75\tsca$, when the shear bands start to fully develop, all four of the norms
start to increase, as small differences between the two simulations build up
over time.

\begin{figure}
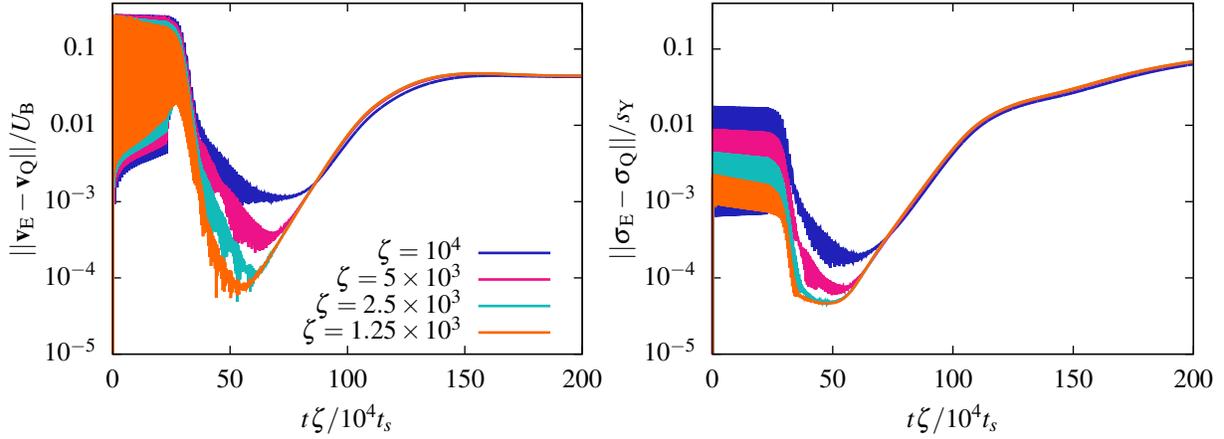

  \begin{center}
    {\footnotesize \include{l2norm2}}
  \end{center}
  \caption{Non-dimensionalized differences between the velocity and stress
  fields in quasi-static and explicit simulations of the rotated line
  configuration, using four different speedup factors $\zeta$, and
  a quasi-static timestep size of $\Delta t = \frac{200\tsca}{\zeta}$. The
  $L_2$ norm defined in the text is used.\label{fig:l2diff2}}
\end{figure}

\begin{figure}
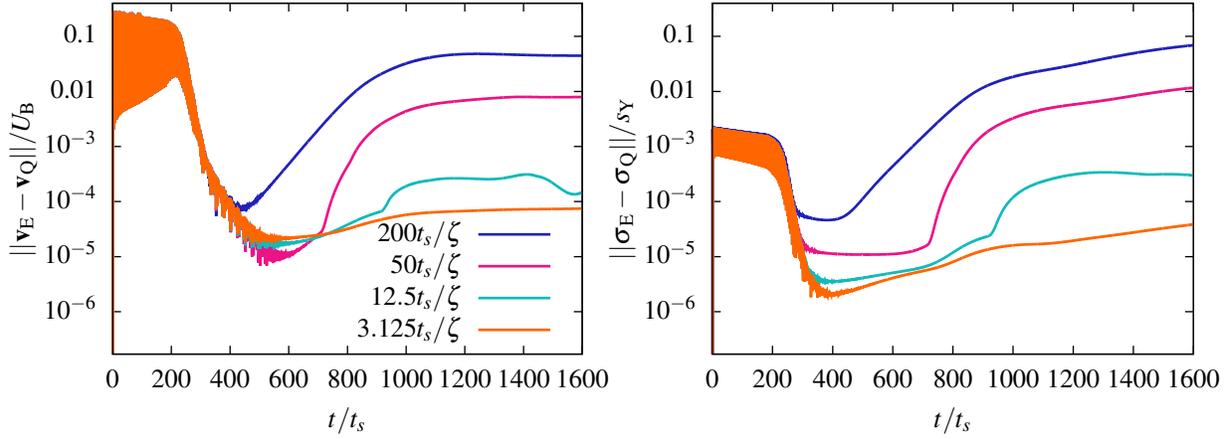

  \begin{center}
    {\footnotesize \include{l2norm3}}
  \end{center}
  \caption{Non-dimensionalized differences between the velocity and stress
  fields in quasi-static and explicit simulations of the rotated line
  configuration, for four different quasi-static timestep sizes, using a
  speedup factor of $\zeta=1.25\times 10^3$. The $L_2$ norm defined in the text
  is used.\label{fig:l2diff3}}
\end{figure}

\begin{figure}
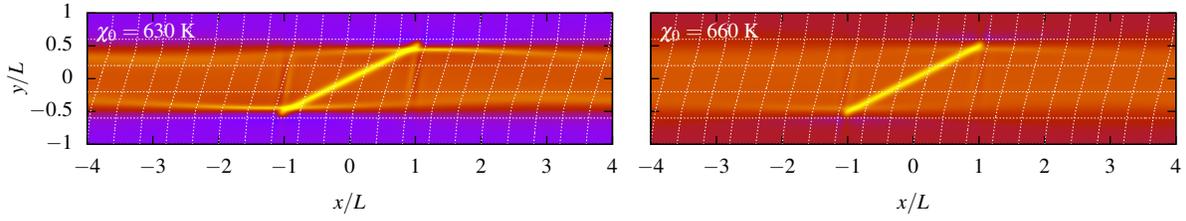

  \begin{center}
    {\scriptsize \include{qline_comp}}
  \end{center}
  \caption{Quasi-static simulation snapshots at $t=200\tsca$ for two higher
  initial values of effective temperature $\chi_0$, using a speedup factor of
  $\zeta=10^4$ and a quasi-static timestep of $\Delta t =
  \frac{200\tsca}{\zeta}$. The color gradient is the same as that used in
  Fig.~\ref{fig:basic_tem}.\label{fig:qline_comp}}
\end{figure}

\begin{figure}
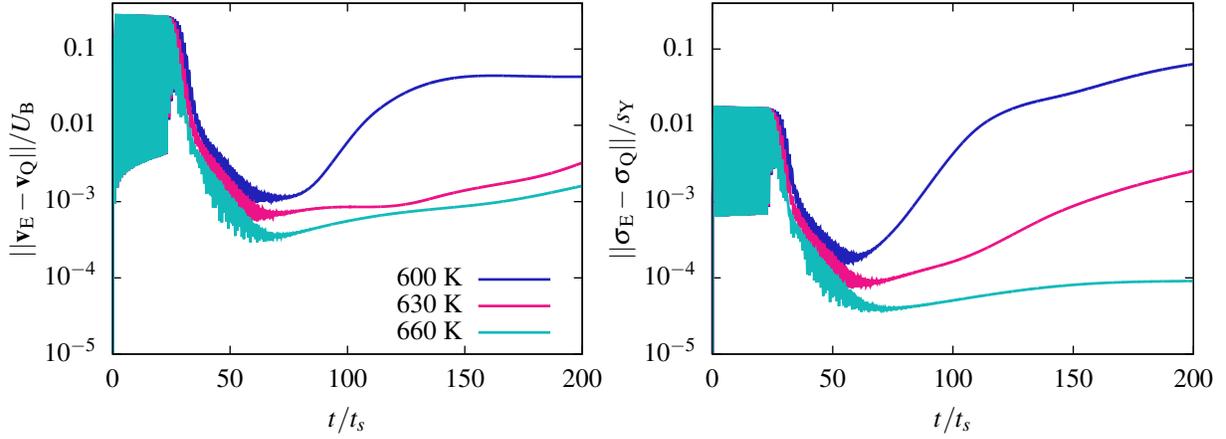

  \begin{center}
    {\footnotesize \include{l2norm4}}
  \end{center}
  \caption{Non-dimensionalized differences between the velocity and stress
  fields in quasi-static and explicit simulations of the rotated line
  configuration, for three different initial background effective temperatures
  $\chi_0$, using a speedup factor of $\zeta=10^4$ and a quasi-static timestep
  of $\Delta t = \frac{200\tsca}{\zeta}$. The $L_2$ norm defined in the text is
  used.\label{fig:l2diff4}}
\end{figure}

Figure~\ref{fig:l2diff2} shows a comparison of the norms for the cases of
$\zeta=10^4, 5 \times 10^3, 2.5\times 10^3, 1.25 \times 10^3$. In the interval
$25\tsca<t<75\tsca$ there is some limited improvement in the agreement between
the methods, but for $t>75\tsca$, all four simulations have near-identical
differences, suggesting that the dominant factor is not $\zeta$ but a
difference in the discretization. Figure~\ref{fig:l2diff3} shows several
simulations for $\zeta=1.25\times10^3$, where the quasi-static timestep size is
reduced by factors of four, sixteen, and 64, substantially improving the
agreement for $t>75\tsca$. However, the agreement for the range
$25\tsca<t<75\tsca$ is unchanged. Comparisons were also carried out using the
original quasi-static timestep and $\zeta=10^4$ for two larger initial
effective temperatures $\chi_0$ in Eq.~\ref{eq:line_chi0}.
Figure~\ref{fig:qline_comp} shows snapshots of these two simulations for
$\chi_0=630\text{~K}$ and $\chi_0=660\text{~K}$ at $t=200\tsca$. For
$\chi_0=630\text{~K}$, there is still some evidence of shear bands nucleating
from $(x,y) \approx (\pm 0.5L,\pm L)$, although they are much weaker than in
Fig.~\ref{fig:qline}, and there is also a large diffuse band of higher
effective temperature in the region $|y|<0.5L$. For $\chi_0=660\text{~K}$ the
thin shear bands are no longer visible, and instead the large diffuse band
dominates. Figure~\ref{fig:l2diff4} shows the differences between the
explicit and quasi-static simulations for the three different values of
$\chi_0$. The simulations for the higher $\chi_0$ agree more closely.

\begin{figure}
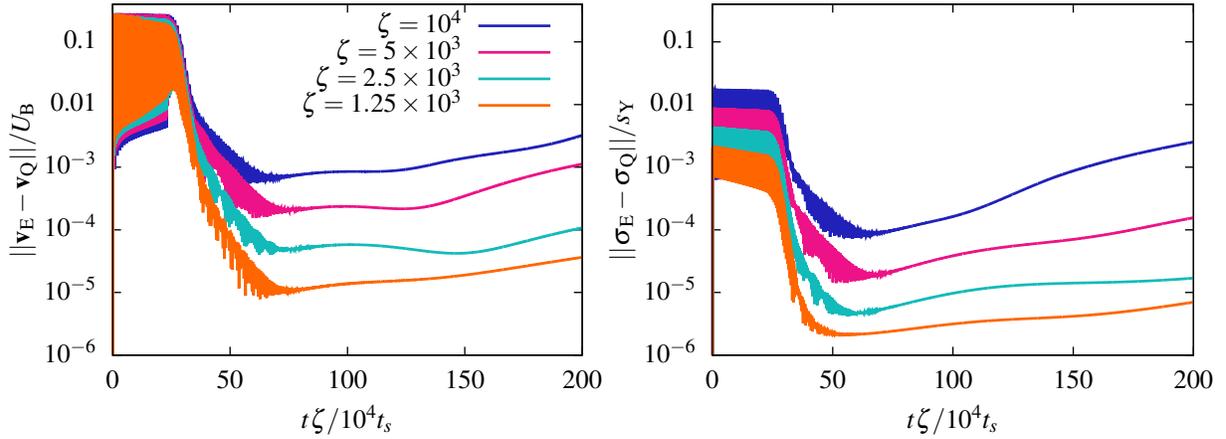

  \begin{center}
    {\footnotesize \include{l2norm5}}
  \end{center}
  \caption{Non-dimensionalized differences between the velocity and stress
  fields in quasi-static and explicit simulations of the rotated line
  configuration for an initial background effective temperature of
  $\chi_0=630\text{~K}$ and a quasi-static timestep of \smash{$\Delta t =
  \frac{3.125\tsca}{\zeta}$}, for four different speedup factors $\zeta$. The
  $L_2$ norm defined in the text is used.\label{fig:l2diff5}}
\end{figure}

Taken together, Figs.~\ref{fig:l2diff2}--\ref{fig:l2diff4} clarify the role of
discretization errors in differences between the two simulations. The largest
differences are caused by the presence of thin shear bands. Since these
features may propagate rapidly across the grid, a relatively small quasi-static
timestep is required in order to properly resolve them. With these results in
mind, we now return to the original question of showing an improvement in
agreement between the two methods as $\zeta$ is reduced. Based on the previous
results, we examine the case of $\chi_0=630\text{~K}$ and a quasi-static
timestep of $\frac{3.125t_s}{\zeta}$, where we expect that the discretization
errors between the two simulations will be small. Figure~\ref{fig:l2diff5}
shows the differences for four values of $\zeta$ and confirms that the
differences are reduced for the entire duration of the simulation as $\zeta$ is
lowered. For $\zeta=1.25\times 10^3$, other than the initial elastic wave
transients, the velocity norm remains below $10^{-4}$ and the stress norm
remains below $10^{-5}$ for the entire duration of the simulation, providing
confidence that the two methods are in very close agreement.

\subsection{Quasi-static simulations of physically realistic timescales}
\begin{figure}
  \begin{center}
    {\scriptsize
    \include{qsin}}
    \setlength{\unitlength}{0.0125bp}
    \vspace{-6mm}
    \begin{picture}(21100,2000)(0,0)
      \footnotesize
      \put(1600,2000){\includegraphics[width=226pt,height=12.5pt]{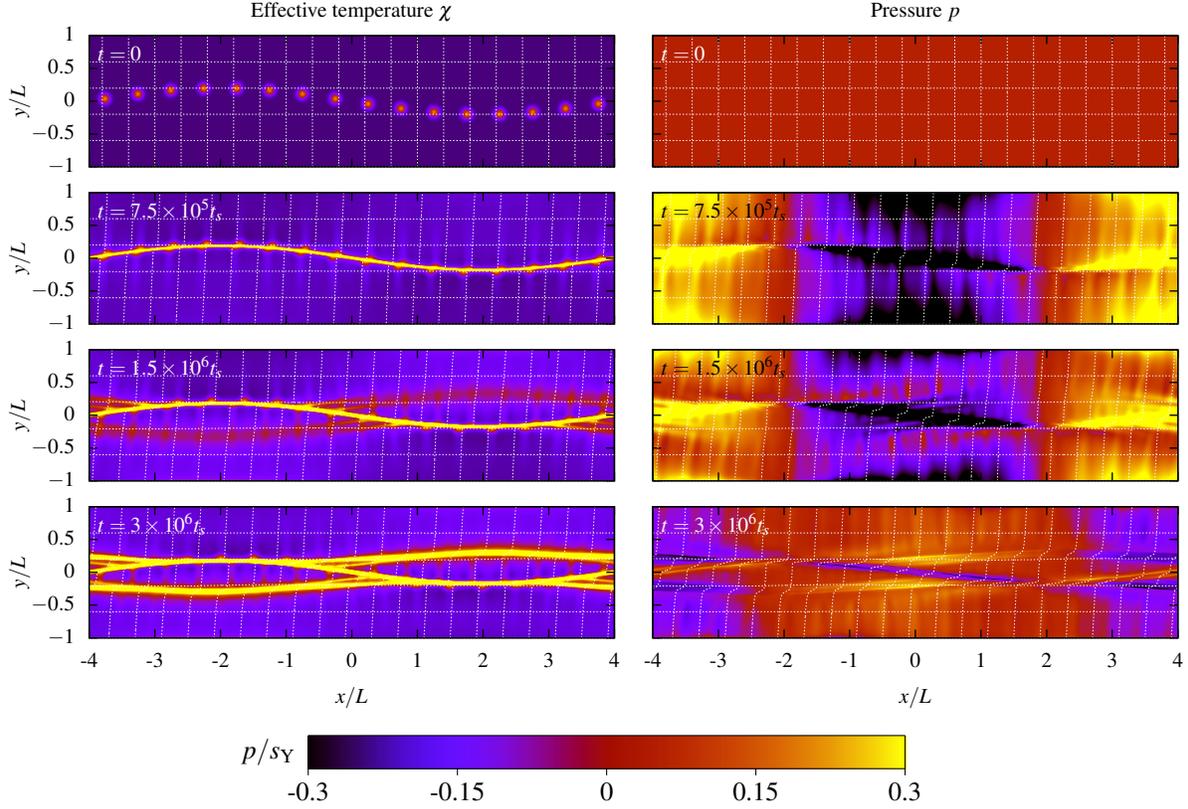}}
      \put(1600,2000){\line(1,0){18000}}
      \put(1600,1800){\line(0,1){1200}}
      \put(1600,3000){\line(1,0){18000}}
      \put(19600,1800){\line(0,1){1200}}
      \put(400,2500){\makebox(0,0)[c]{$p/\sY$}}
      \put(6100,1800){\line(0,1){200}}
      \put(10600,1800){\line(0,1){200}}
      \put(15100,1800){\line(0,1){200}}
      \put(1600,1300){\makebox(0,0)[c]{-0.3}}
      \put(6100,1300){\makebox(0,0)[c]{-0.15}}
      \put(10600,1300){\makebox(0,0)[c]{0}}
      \put(15100,1300){\makebox(0,0)[c]{0.15}}
      \put(19600,1300){\makebox(0,0)[c]{0.3}}
    \end{picture}
  \end{center}
  \caption{Snapshots of effective temperature $\chi$ and pressure $p$ for a
  quasi-static simulation with $\zeta=1$. The color gradient for the effective
  temperature is the same as that used in
  Fig.~\ref{fig:basic_tem}.\label{fig:sine_temp}}
\end{figure}

For realistic strain rates, the explicit simulation method becomes
prohibitively expensive but the quasi-static simulation method remains
feasible. Here, we demonstrate this capability by simulating an example using
$\zeta=1$. In the previous examples considered, there is a strong tendency for
shear bands to form horizontally, even when a non-horizontal feature is
present. Here, we consider a case specifically aimed at forcing a curved shear
band to form. Sixteen positions
\smash{$\vx_k=(\frac{kL}{2}+\frac{L}{4},-\frac{L}{5}\sin
(\frac{\pi}{4}(\frac{k}{2}+\frac{1}{4})))$} for $k=-8, -7, \ldots, 7$ in the
shape of a sine wave are introduced, and the effective temperature is
initialized to be
\begin{equation}
  \chi(\vx,t) = 620\text{~K} + (180\text{~K})\exp\left(-\frac{20^2 \min_k \{ |\vx - \vx_k|^2 \}}{2L^2}\right).
\end{equation}
Figure~\ref{fig:sine_temp} shows a sequence of snapshots of effective
temperature and pressure, using a quasi-static timestep of $\Delta t =
100\tsca$. By $t=7.5\times 10^5\tsca$, a sinusoidal shear band has formed that
links together the initial regions of higher $\chi$. Shearing along
this sinusoidal band causes material to be pushed toward the region of
$(x,y)=(0,4L)$ and be pulled away from $(x,y)=(0,0)$, resulting in large
positive and negative pressures respectively at these locations. By
$t=1.5\times 10^6\tsca$, a further pair of shear bands start to emerge, which
become fully developed by $t=3\times 10^6 \tsca$. The additional shear bands
allow the material to shear more easily and the pressure is reduced.

\section{Free boundary simulations}
\label{sec:free}
The two-dimensional shearing simulations that have been considered in the
previous sections employ simple boundary conditions where the velocity is
prescribed on all of the physical boundaries. This leads to Dirichlet boundary
conditions for the elliptic problem in the projection step, which are
straightforward to implement. In this section, we extend the method to handle
objects with moving boundaries to make it applicable to more general solid
mechanics problems. We focus on the application of a traction-free condition
$\bsig \cdot \nor = \vec{0}$ at a boundary where $\nor$ is an outward-pointing
normal vector.

There is again a close parallel with the fluid projection method,
where conditions such as $\vv\cdot \nor=0$ are frequently applied to enforce
no normal flow across an impermeable boundary. At the
end of a timestep, one wishes to enforce that $\nor \cdot \vv_{n+1}=0$.
Taking the inner product of Eq.~\ref{eq:fproj2} with $\nor$ yields
\begin{equation}
  \label{eq:f_nbc}
\frac{\rho\nor \cdot \vv_*}{\Dt} = \nor \cdot \nabla p_{n+1},
\end{equation}
which is a Neumann condition in the elliptic problem for $p_{n+1}$. In a case
when all boundaries in a computation are of the form $\vv \cdot \nor=0$, so
that the elliptic problem for pressure only employs Neumann conditions, the
pressure is only determined up to an additive constant.

Analogous steps can be taken for quasi-static elastoplasticity to apply the
traction-free condition at the end of a time step, so that $\nor \cdot
\bsig_{n+1}=\vec{0}$. Taking the inner product of Eq.~\ref{eq:sproj2} with
$\nor$ yields
\begin{equation}
  \label{eq:s_nbc}
  - \frac{\nor \cdot \bsig_*}{\Dt} = \nor \cdot \tC : \tD_{n+1}.
\end{equation}
This is similar to a Neumann condition: it enforces two conditions on the
gradients of the components of $\vv$, although there is also a coupling. If a
problem is considered where traction-free conditions are applied everywhere,
such as for an object freely floating in space, then the velocity will only be
determined up to an additive vector constant. This is physically reasonable
since the original system of equations, Eqs.~\ref{eq:constit} and
\ref{eq:quasis}, does not have any preferred velocity. Pinning the velocity at
a single point in a freely floating body is enough to set the additive constant
and determine the entire velocity field.

\subsection{Boundary representation}
To track the free boundary of an object we make use of the level set
method~\cite{osher88}, whereby an auxiliary function $\phi(\vx,t)$ is
introduced and is initialized to be the signed distance to the boundary, with
the convention that $\phi(\vx,t)<0$ inside the simulated object and
$\phi(\vx,t)>0$ outside the object. The level set method is well-suited to an
Eulerian framework, since the function $\phi$ can be discretized on the same
Cartesian grid as other simulation fields. It provides an implicit
representation of the boundary as the zero contour, $\phi(\vx,t)=0$. The method
is widely used in fluid mechanics, since it can easily handle large stretches
and topology changes in the boundary.

In principle, given an interface moving according to a globally defined
velocity field $\vv(\vx,t)$, the function $\phi(\vx,t)$ can be updated by
using the transport equation
\begin{equation}
  \frac{\p \phi}{\p t} + (\vv \cdot \nabla) \phi =0.
  \label{eq:basicls}
\end{equation}
However in practice this causes a number of numerical difficulties: while the
zero contour of $\phi$ will remain at the interface, the function $\phi$ may no
longer be a signed distance function to the interface. In addition, for the
current problem the simulation fields only exist on one side of the level set,
inside the object where $\phi(\vx,t)\le 0$, and it is therefore not clear what
value of $\vv$ to use in Eq.~\ref{eq:basicls}.

These issues have been extensively studied over the past two decades and for a
full treatment the reader should consult the books by Sethian~\cite{sethian}
and Osher~\cite{osher}. The signed distance property can be maintained by
periodically reinitializing $\phi$, such as by using a PDE-based
approach~\cite{sussman94} or by a fast marching method~\cite{sethian}. Given
fields defined inside a body, the level set function can also be used to
extrapolate those fields along rays normal to the interface~\cite{aslam04},
which can be used to apply boundary conditions, or to construct a globally
defined $\vv$ in order to apply Eq.~\ref{eq:basicls}. For computational
efficiency, the level set function only needs to be stored on a narrow band of
grid points surrounding $\phi(\vx,t)=0$.

For the examples considered here, we make use of the specific level set
implementation that was previously developed for simulating elastoplastic
dynamics~\cite{rycroft12}. The method employs a narrow-banded level set for
efficiency, and makes use of a combination of a second-order fast marching
method and the modified Newton--Raphson algorithm of Chopp~\cite{chopp01}. It
continually keeps the level set function close to a signed distance function,
without the need for specific reinitialization operations. The simulation
fields can be linearly extrapolated. We also make use of routines first
discussed in Kamrin {\it et al.}~\cite{kamrin12} that can linearly extrapolate
fields stored on a grid staggered with respect to the level set field. In the
examples that follow, the results are not strongly dependent on the specifics
of the level set implementation and we therefore refer the reader to these
previous papers for more details.

\subsection{Numerical framework}
The examples considered here make use of a non-periodic grid of $M \times N$
square cells. As in the previous sections, the stress and effective temperature
are stored at cell centers, while the velocity field and reference map are
stored at cell corners. The level set field is stored at cell centers, and is
initialized to represent a shape that is attached to the boundary at one or
more locations, where the conditions
\begin{equation}
  \vv(\vx,t) = \vec{0}, \qquad \bxi(\vx,t)=\vx
  \label{eq:lsbc}
\end{equation}
are used. The simulation fields are only updated at grid points that are inside
the body. A cell center \smash{$(i+\frac{1}{2},j+\frac{1}{2})$} is defined as
inside the body if the level set field satisfies $\phi_{i+1/2,j+1/2}<0$. A cell
corner $(i,j)$ is defined as inside the body if the bilinear interpolation of
the level set field
\begin{equation}
\phi^\prime_{i,j}=\frac{\phi_{i-1/2,j-1/2}+\phi_{i+1/2,j-1/2}+\phi_{i-1/2,j+1/2}+\phi_{i+1/2,j+1/2}}{4}
\end{equation}
satisfies $\phi^\prime_{i,j}<0$. As described above, given a particular
simulation field $f_{i,j}$ defined at grid points inside the body, linearly
extrapolated values $f^\text{ex}_{i,j}$ at points outside the body can be
calculated. Prior to performing a simulation, all fields are extrapolated.

To carry out a timestep of $\Delta t$ in the free boundary simulations, the
following procedure is used for both the explicit and quasi-static methods:
\begin{enumerate}
  \item Move the level set according to the velocity field.
  \item Using the new level set values, update which points are inside the
    body. Initialize the simulation fields at any new grid points inside the body
    to be equal to the extrapolated values.
  \item Calculate the finite-difference update using either the explicit method
    described in Subsec.~\ref{sub:directsim} or the quasi-static method described
    in Subsec.~\ref{sub:quasisim}, taking into account boundary conditions at the
    free boundary.\label{ite:fd}
  \item Extrapolate all fields.
  \item Enforce the boundary conditions of Eq.~\ref{eq:lsbc}.
\end{enumerate}
Step~\ref{ite:fd} requires additional consideration for both the explicit and
quasi-static methods. In the explicit simulation, the velocity $\vv$, reference
map field $\bxi$, and effective temperature $\chi$ are unconstrained at the
free boundary. Hence, when a finite-difference calculation references any
exterior point, it makes use of the available extrapolated value. The
simulation only ever makes use of the exterior points that are directly
adjacent to interior points. If an ENO calculation would reference an exterior
point that is two points away from the interior, then the simulation falls back
on a first-order upwinded derivative.

The stress tensor $\bsig$ must be handled differently in order to apply the
traction-free boundary condition $\bsig \cdot \nor = \vec{0}$. When calculating
the advective derivatives, the simulation makes use of the same procedure as
described in previous work~\cite{rycroft12}, where a modified extrapolated
value is calculated so that the linear interpolation of the stress field will
satisfy the traction-free condition at the precise location of the zero level
set. In addition to this, a similar procedure must be introduced to handle the
boundary condition when evaluating the stresses in Eqs.~\ref{eq:direct_start}
and \ref{eq:vel2} since the velocity field is staggered with respect to the
stress field. Consider updating the velocity at a grid location $(i,j)$ and
suppose that the cell center \smash{$(i+\frac{1}{2},j+\frac{1}{2})$} is an
exterior point. Then
\begin{equation}
  \alpha= \frac{\phi^\prime_{i,j}}{\phi^\prime_{i,j}-\phi_{i+1/2,j+1/2}}
\end{equation}
represents the position along the diagonal line from $(i,j)$ to
\smash{$(i+\frac{1}{2},j+\frac{1}{2})$} where the zero level set intersects. At
this intersected position, an interpolated stress is calculated as
\begin{equation}
  \bsig_P=\frac{(1+\alpha)\bsig_{i+1/2,j+1/2}+(1-\alpha)\bsig_{i-1/2,j-1/2}}{2}
\end{equation}
and a normal vector is calculated as the gradient of the bilinear interpolation
of $\phi$. Following previous work~\cite{rycroft12} a new $\bsig^\prime_P$
is then constructed where the normal--normal and normal--tangential stress
components are projected to zero. Finally, a modified extrapolated value at
\smash{$(i+\frac{1}{2},j+\frac{1}{2})$} is calculated as
\begin{equation}
  \bsig^\prime_{i+1/2,j+1/2} = \frac{2\bsig^\prime_P-(1-\alpha)\bsig_{i-1/2,j-1/2}}{1+\alpha},
\end{equation}
which is then used in the finite-difference calculation of Eqs.~\ref{eq:direct_start}
and \ref{eq:vel2}.

\subsection{Boundary implementation in the projection step}
\label{sub:quasi_boundary}
The projection step in the quasi-static method must also be modified to take
into account the free boundary. The velocity fields must only be solved at grid
points within the body. At these points, the linear system is constructed in
the same manner as previously, using the discretization of
Eqs.~\ref{eq:double_multi1} and \ref{eq:double_multi2}. The discretization will
also reference exterior grid points that are either orthogonally or
diagonally adjacent to an interior point---we refer to this set of outside
points as neighboring points.

At the neighboring points, we also solve for the velocity in the linear system,
and calculate values that are consistent with the boundary condition in
Eq.~\ref{eq:s_nbc}, which is
\[
  \frac{1}{\Delta t} \hat{\vec{n}} \cdot
\left(
\begin{array}{cc}
  -p_*-q_* + s_* & \tau_* \\
  \tau_* & -p_*-q_* -s_* \\
\end{array}
\right)
\]
\begin{equation}
  =\hat{\vec{n}} \cdot
\left(
\begin{array}{cc}
  -K'(u_x+v_y) - \mu(u_x-v_y) & -\mu(u_y+v_x) \\
  -\mu(u_y+v_x) & -K'(u_x+v_y) +\mu(u_x-v_y) \\
\end{array}
\right)\label{eq:s_nbc_comp}
\end{equation}
when expressed in terms of the simulation fields. Applying this condition is
similar to extrapolation~\cite{aslam04,sethian,rycroft12}, in that the
velocities at the neighboring points are normally extended from the interior
points in a manner that satisfies Eq.~\ref{eq:s_nbc_comp}.

\begin{figure}
  \setlength{\unitlength}{0.86bp}
  \begin{center}
    {\footnotesize
    \begin{picture}(500,160)(0,0)
      \put(0,20){\includegraphics[scale=0.86]{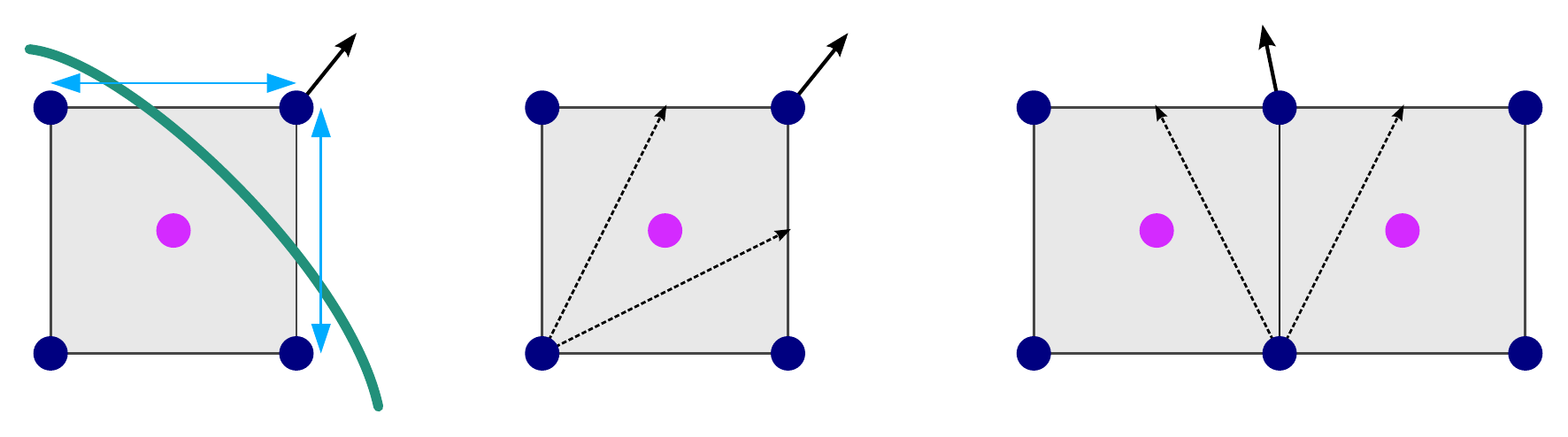}}
      \put(49,0){(a)}
      \put(209,0){(b)}
      \put(409,0){(c)}
      \put(4,32){\textcolor{dblue}{$(\vv_{dl})$}}
      \put(6,112){\textcolor{dblue}{$\vv_l$}}
      \put(86,32){\textcolor{dblue}{$\vv_d$}}
      \put(86,112){\textcolor{dblue}{$\vv$}}
      \put(46,72){\textcolor{dmag}{$\vec{b}$}}
      \put(206,72){\textcolor{dmag}{$\vec{b}$}}
      \put(366,72){\textcolor{dmag}{$\vec{b}_l$}}
      \put(446,72){\textcolor{dmag}{$\vec{b}_r$}}
      \put(98,18){\textcolor{jade}{$\phi(\vx,t)=0$}}
      \put(52,142){\small \textcolor{lblue}{$\frac{\p \vv}{\p x}$}}
      \put(107,82){\small \textcolor{lblue}{$\frac{\p \vv}{\p y}$}}
      \put(166,32){\textcolor{dblue}{$\vv_{dl}$}}
      \put(166,112){\textcolor{dblue}{$\vv_l$}}
      \put(246,32){\textcolor{dblue}{$\vv_d$}}
      \put(246,112){\textcolor{dblue}{$\vv$}}
      \put(326,32){\textcolor{dblue}{$\vv_{dl}$}}
      \put(326,112){\textcolor{dblue}{$\vv_l$}}
      \put(406,32){\textcolor{dblue}{$\vv_d$}}
      \put(406,112){\textcolor{dblue}{$\vv$}}
      \put(486,32){\textcolor{dblue}{$\vv_{dr}$}}
      \put(486,112){\textcolor{dblue}{$\vv_r$}}
      \put(113,154){$\nor$}
      \put(273,154){$\nor$}
      \put(408,157){$\nor$}
    \end{picture}}
  \end{center}
  \vspace{1.5mm}
  \caption{(a) Schematic of the basic procedure to set the velocity $\vv$ at an
  exterior point to be consistent with the traction-free boundary condition.
  The boundary condition involves the source term $\vec{b}$ at the cell center,
  the normal vector $\nor$ at the exterior point, and the derivatives $\p_x
  \vv$ and $\p_y \vv$, which can be approximated by $(\vv-\vv_l)/h$ and
  $(\vv-\vv_d)/h$ respectively. (b,~c)~Representative diagrams showing the
  two types of boundary conditions, which are used if $\nor$ lies within the
  range of angles shown by the dashed arrows.\label{fig:extrap_diag}}
\end{figure}

To illustrate this procedure, consider the basic example shown in
Fig.~\ref{fig:extrap_diag}(a), where the velocity at the neighboring point
$\vv$ can be expressed in terms of the velocities at the interior points
$\vv_d$ and $\vv_l$. One-sided first derivatives of $\vv$ are given by
\begin{equation}
\frac{\p \vv}{\p x} = \frac{\vv-\vv_l}{h}, \qquad \frac{\p \vv}{\p y} = \frac{\vv-\vv_d}{h}.
\end{equation}
A normal vector $\nor$ is calculated at $\vv$. A source term $\vec{b}=-
\frac{\nor \cdot \bsig_*}{\Dt}$ is then calculated at the center of the square.
If the two matrices
\begin{equation}
H(\nor) = \frac{1}{h} \left(
\begin{array}{cc}
  -(K'+\mu) n_x & -\mu n_y \\
  (\mu-K') n_y & -\mu n_x
\end{array}
\right),
\qquad
V(\nor) = \frac{1}{h} \left(
\begin{array}{cc}
  -\mu n_y & (\mu-K') n_x \\
  -\mu n_x & -(K'+\mu) n_y
\end{array}
\right)
\end{equation}
are introduced, then Eq.~\ref{eq:s_nbc_comp} can be implemented as
\begin{equation}
H(\nor) (\vv-\vv_l) + V(\nor) (\vv-\vv_d) = \vec{b}.
\end{equation}

From Fig.~\ref{fig:extrap_diag}(a) it can be seen that there is some freedom in
choosing the precise formula for $\vv$. For example, the $x$-derivative could
be also obtained using $\p\vv/\p x=(\vv_{d} - \vv_{dl})/h$. In our numerical
tests, we found that the best results were achieved when extension formulae
made use of a combination of the available velocities that closely matched with
the direction of the normal vector. We therefore made use of two different
types of numerical stencils depending on whether the normal vector pointed
diagonally or orthogonally. The stencils are chosen in such a way that their
values change continuously as the angle of the normal vector is varied.

The first stencil type is shown in Fig.~\ref{fig:extrap_diag}(b) and is
illustrated for the case when the normal vector points diagonally up-right so
that $2\hat{n}_x>\hat{n}_y$ and $2\hat{n}_y>\hat{n}_x$. A variable $\beta$ is
defined as
\begin{equation}
\beta=\left\{
\begin{array}{ll}
  \dfrac{\hat{n}_y}{2\hat{n}_x} & \qquad \text{if $\hat{n}_y>\hat{n}_x$,} \\
  1-\dfrac{\hat{n}_x}{2\hat{n}_y} & \qquad \text{if $\hat{n}_x\ge\hat{n}_y$,}
\end{array}
\right.
\end{equation}
so that it continuously varies from 0 to 1 over the range of normal vectors
considered. If $\alpha=1-\beta$, then the boundary condition is implemented as
\begin{eqnarray}
H(\nor) \left[ \beta (\vv - \vv_l) + \alpha(\vv_d -\vv_{dl})\right]
+ V(\nor) \left[ \alpha (\vv-\vv_d) + \beta(\vv_l - \vv_{dl})\right]
&& \nonumber \\
+ 8\beta \alpha \left(\alpha V(\nor) +\beta H(\nor)\right) (\vv + \vv_{dl} - \vv_l - \vv_d)&=&
\vec{b},
\end{eqnarray}
where the source term $\vec{b}$ is calculated at the center of the grid cell.
This formulation therefore smoothly transitions from calculating derivatives on
the bottom and right cell edges when $\beta=0$, to calculating derivatives on
top and left cell edges when $\beta=1$. The third term on the left hand side of
the equation amplifies the diagonal terms when the normal is close to the
diagonal.

The second stencil type is shown in Fig.~\ref{fig:extrap_diag}(c) and is
illustrated for cases where the normal vector points upward, so that
$\hat{n}_y\ge2|\hat{n}_x|$. In this case, the variable is defined as
\begin{equation}
  \beta=\frac{1}{2} + \frac{\hat{n}_x}{\hat{n}_y}
\end{equation}
so that it varies from 0 to 1 over the range of normal vectors considered.
If $\alpha=1-\beta$, the boundary condition is implemented as
\begin{equation}
  V(\nor) \left[\vv - \vv_d\right] + H(\nor) \left[ \beta (\vv_{dr} -\vv_d) + \alpha(\vv_d-\vv_{dl}) \right] = \beta \vec{b}_r + \alpha \vec{b}_l,
\end{equation}
where $\vec{b}_l$ and $\vec{b}_r$ are the source terms on the left and right
grid cells. By applying flips in the $x$ and $y$ axes, the two stencils shown
in Figs.~\ref{fig:extrap_diag}(b) and \ref{fig:extrap_diag}(c) can be extended
to handle all other directions of normal vector. As the normal vector changes,
the stencil entries and the source terms that are used all vary continuously,
and there are no discontinuous jumps between the different cases.

\subsection{Quasi-static loading and unloading of a bar}
\label{sub:qloadu}
The first free boundary example makes use of a horizontal bar where the right
end is fixed to a wall. At the left end of the bar, a load is incrementally
applied on a quasi-static timescale, and is then incrementally removed. The
load is applied in a diagonal direction so that the bar is both stretched and
deformed downward, and the magnitude of the load is large enough to cause a
substantial amount of plastic deformation around the loading region. This leads
to a complex deformation of the bar, which makes for a good numerical test of
the method. By using the reference map field $\bxi(\vx,t)$, we also demonstrate
the calculation of strain in a fully Eulerian simulation, and we examine the
interplay between deviatoric and volumetric strain.

The example uses the domain $-2L \le x \le 2L, -L \le y \le L$ with a $512
\times 256$ grid. The level set is initialized to represent a horizontal bar in
the region $x>-1.65L, |y|<0.65L$ with rounded corners of radius $0.3L$, due to
the difficulties of accurately representing sharp corners using the level set
method. The bar is fixed to the boundary at $x=2L$, and the initial effective
temperature in the bar is 620~K. The simulation lasts for $10^6\tsca$, which is
$4.05\text{~s}$ for the nominal length scale of $L=1\text{~cm}$. Quasi-static
timesteps of size $1250\tsca$ are used. The load position is given by
$\vx_F(t)$ with initial condition $\vx_F(0)=(-L,0)$. The load moves with the
body according to
\begin{equation}
  \frac{d\vx_F}{dt} = \vv(\vx_F,t).
\end{equation}
This equation is implemented using the Euler timestep, and the term
$\vv(\vx_F,t)$ is calculated using bicubic interpolation of the velocity field.
The load is applied as a body force $\vec{F}(\vx,t)$ in the projection step, as
an additional source term on the right hand side of Eqs.~\ref{eq:double_multi1}
and \ref{eq:double_multi2}. The time dependence of the applied load is given by
the function
\begin{equation}
F_T(t) = \left\{
\begin{array}{ll}
  \frac{t}{\tsca} & \qquad \text{for $0\le t < 4\times 10^5 \tsca$,} \\
  8\times 10^5 - \frac{t}{\tsca} & \qquad \text{for $4\times 10^5 \tsca \le t < 8\times 10^5 \tsca$,} \\
  0 & \qquad \text{for $8\times 10^5 \tsca \le t \le 10^6 \tsca$}
\end{array}
\right.
\label{eq:load_time}
\end{equation}
so that the bar is incrementally loaded up to $t=4\times 10^5\tsca$ and then
incrementally unloaded up to $t=8\times 10^5\tsca$. The spatial dependence of
the applied load is given by
\begin{equation}
F_R(r) = \left\{
\begin{array}{ll}
  1 + \cos \frac{\pi r}{r_F} & \qquad \text{for $r<r_F$,} \\
  0 & \qquad \text{for $r\ge r_F$,}
\end{array}
\right.
\label{eq:load_space}
\end{equation}
so that it is applied over a circle of radius $r_F=0.25L$. The force is then
given in terms of these two functions as
\begin{equation}
\vec{F}(\vx,t) = - \left(
\begin{array}{c}
  12\psi \\ \psi
\end{array}
\right) F_R(|\vx-\vx_F|) F_T(t),
\label{eq:load}
\end{equation}
where $\psi=4.625\times10^{-6} \sY/L$.

\begin{figure}
  \begin{center}
    \setlength{\unitlength}{0.0125bp}
    \begin{picture}(35500,35500)(0,0)
      \put(0,17000){\parbox{15cm}{\scriptsize \include{bdry1a}}}
      \setlength{\unitlength}{0.0125bp}
      \put(29000,8290){
	\begin{picture}(2000,21100)(0,0)
	  \footnotesize
	  \put(0,1600){\includegraphics[width=12.5pt,height=226pt]{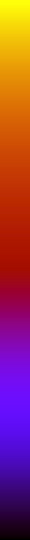}}
	  \put(0,1600){\line(0,1){18000}}
	  \put(0,1600){\line(1,0){1200}}
	  \put(1000,1600){\line(0,1){18000}}
	  \put(0,19600){\line(1,0){1200}}
	  \put(500,400){\makebox(0,0)[c]{$p/\sY$}}
	  \put(1000,6100){\line(1,0){200}}
	  \put(1000,10600){\line(1,0){200}}
	  \put(1000,15100){\line(1,0){200}}
	  \put(1300,1600){\makebox(0,0)[l]{-1}}
	  \put(1300,6100){\makebox(0,0)[l]{-0.5}}
	  \put(1300,10600){\makebox(0,0)[l]{0}}
	  \put(1300,15100){\makebox(0,0)[l]{0.5}}
	  \put(1300,19600){\makebox(0,0)[l]{1}}
	\end{picture}}
      \put(33000,8290){
	\begin{picture}(2000,21100)(0,0)
	  \footnotesize
	  \put(0,1600){\includegraphics[width=12.5pt,height=226pt]{colorchart_v}}
	  \put(0,1600){\line(0,1){18000}}
	  \put(0,1600){\line(1,0){1200}}
	  \put(1000,1600){\line(0,1){18000}}
	  \put(0,19600){\line(1,0){1200}}
	  \put(500,400){\makebox(0,0)[c]{$|\bsig_0|/\sY$}}
	  \put(1000,6100){\line(1,0){200}}
	  \put(1000,10600){\line(1,0){200}}
	  \put(1000,15100){\line(1,0){200}}
	  \put(1300,1600){\makebox(0,0)[l]{0}}
	  \put(1300,6100){\makebox(0,0)[l]{0.3}}
	  \put(1300,10600){\makebox(0,0)[l]{0.6}}
	  \put(1300,15100){\makebox(0,0)[l]{0.9}}
	  \put(1300,19600){\makebox(0,0)[l]{1.2}}
	\end{picture}}
    \end{picture}
  \end{center}
  \caption{Plots of pressure $p$ (left) and deviatoric stress $|\bsig_0|$
  (right) at five time points of the stretched bar simulation. The boundary of
  the bar is shown as the solid white line obtained as the zero contour of
  level set function $\phi$. The thin dashed white lines are the contours of
  the components of the reference map $\bxi$ and show how the material is
  deformed. The dashed cyan circle shows the region where the bar is
  loaded.\label{fig:bdry1_devtem}}
\end{figure}

Figure~\ref{fig:bdry1_devtem} shows snapshots of the pressure and deviatoric
stress in the simulation. As the bar is loaded up to $t=4\times10^5\tsca$,
negative pressures build up in the bar, apart from a small region to the left
of the applied load, where a positive pressure emerges. By $t=4\times
10^5\tsca$ the deviatoric stress has exceeded $\sY$ in some areas, leading to
plastic deformation. After the bar has been unloaded at $t=8\times 10^5\tsca$,
some residual pressure and shear stress is visible as a result of the plastic
deformation. While not shown, the simulation fields remain static over the
interval $8\times 10^5\tsca < t \le 10^6\tsca$.

\begin{figure}
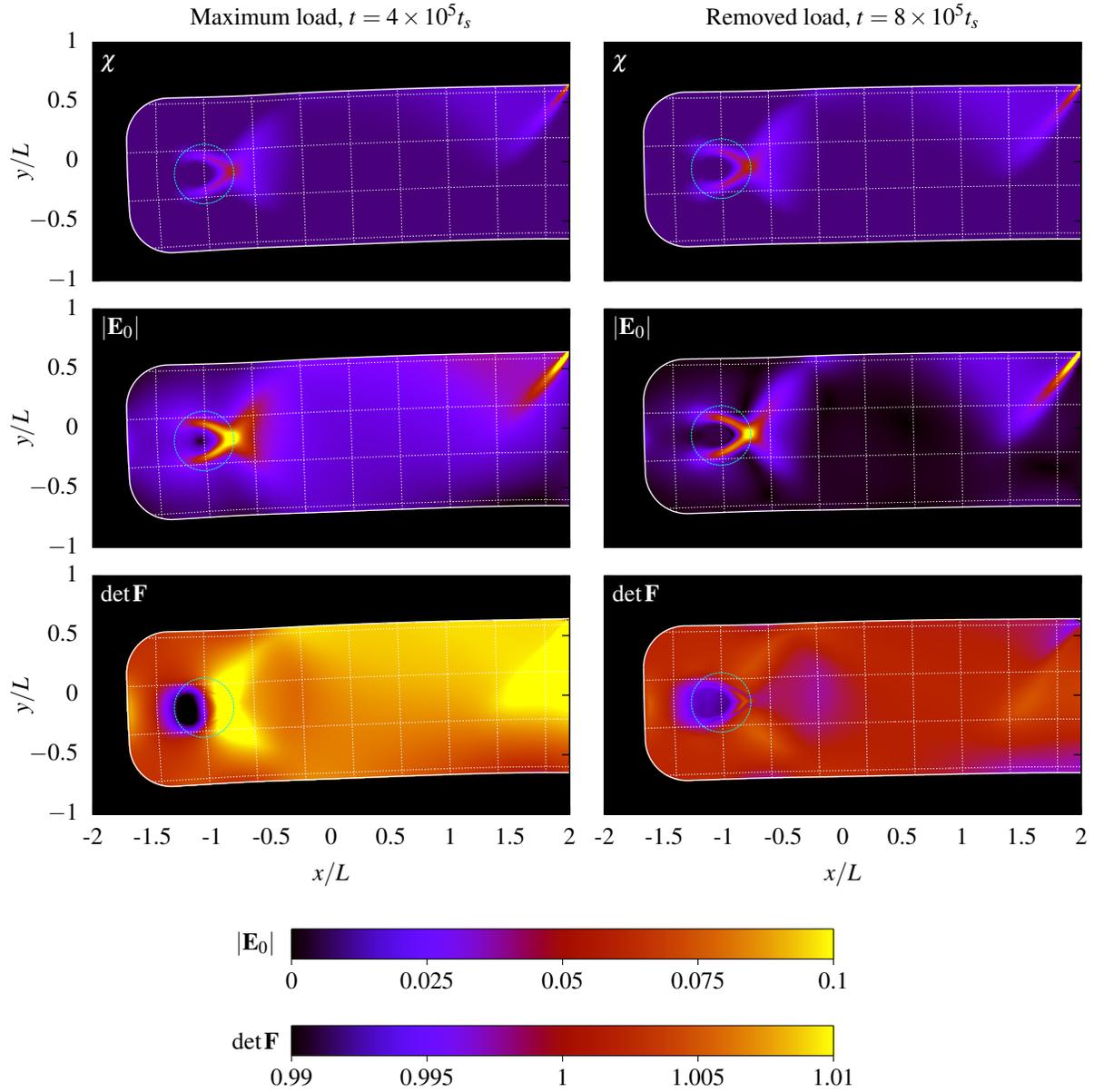

  \begin{center}
    {\footnotesize
    \include{bdry1b}}
    \vspace{-1mm}
    \setlength{\unitlength}{0.0125bp}
    \begin{picture}(21100,2000)(0,0)
      \footnotesize
      \put(1600,2000){\includegraphics[width=226pt,height=12.5pt]{colorchart}}
      \put(1600,2000){\line(1,0){18000}}
      \put(1600,1800){\line(0,1){1200}}
      \put(1600,3000){\line(1,0){18000}}
      \put(19600,1800){\line(0,1){1200}}
      \put(400,2500){\makebox(0,0)[c]{$|\vec{E}_0|$}}
      \put(6100,1800){\line(0,1){200}}
      \put(10600,1800){\line(0,1){200}}
      \put(15100,1800){\line(0,1){200}}
      \put(1600,1300){\makebox(0,0)[c]{0}}
      \put(6100,1300){\makebox(0,0)[c]{0.025}}
      \put(10600,1300){\makebox(0,0)[c]{0.05}}
      \put(15100,1300){\makebox(0,0)[c]{0.075}}
      \put(19600,1300){\makebox(0,0)[c]{0.1}}
    \end{picture} \\ \vspace{5mm}
    \begin{picture}(21100,2000)(0,0)
      \footnotesize
      \put(1600,2000){\includegraphics[width=226pt,height=12.5pt]{colorchart}}
      \put(1600,2000){\line(1,0){18000}}
      \put(1600,1800){\line(0,1){1200}}
      \put(1600,3000){\line(1,0){18000}}
      \put(19600,1800){\line(0,1){1200}}
      \put(400,2500){\makebox(0,0)[c]{$\det \vec{F}$}}
      \put(6100,1800){\line(0,1){200}}
      \put(10600,1800){\line(0,1){200}}
      \put(15100,1800){\line(0,1){200}}
      \put(1600,1300){\makebox(0,0)[c]{0.99}}
      \put(6100,1300){\makebox(0,0)[c]{0.995}}
      \put(10600,1300){\makebox(0,0)[c]{1}}
      \put(15100,1300){\makebox(0,0)[c]{1.005}}
      \put(19600,1300){\makebox(0,0)[c]{1.01}}
    \end{picture}
  \end{center}
  \caption{Plots of effective temperature $\chi$ (top), deviatoric strain
  $|\vec{E_0}|$ (middle), and volume ratio $\det \vec{F}$ (bottom) in the
  stretched bar simulation at the time of maximum load (left) and at the time
  when the load has been removed (right). The boundary of the bar is shown as
  the solid white line obtained as the zero contour of level set function
  $\phi$. The thin dashed white lines are the contours of the components of the
  reference map $\bxi$ and show how the material is deformed. The dashed cyan
  circle shows the region where the bar is loaded. The effective temperature
  plots use the same scale as Fig.~\ref{fig:basic_tem}, and scales for strain
  plots are shown.\label{fig:bdry1_strain}}
\end{figure}

The top two plots in Fig.~\ref{fig:bdry1_strain} show the effective temperature
at the time of maximum load, and at the time when the load is removed. As would
be expected from the regions of high deviatoric stress at $t=4\times10^5\tsca$,
regions of increased $\chi$ are visible around the loading region, and also at
the top right corner, where a small shear band forms. While the bulk of the
increased $\chi$ occurs during the period of increasing load, a small increase
in $\chi$ is also visible during the period of decreasing load---this is
expected since the plastic deformation will not immediately cease when the load
starts to decrease.

Figure~\ref{fig:bdry1_strain} also shows plots of the deviatoric strain
measured in terms of the $|\vec{E}_0|$, and the volume ratio $\det \vec{F}$,
which are computed using the reference map field $\bxi(\vx,t)$. We use $\det
\vec{F}-1$ to measure the volumetric strain. As expected, there is a high
correlation between the deviatoric strain and the regions of higher $\chi$,
since $\chi$ increases in regions where the material has yielded plastically,
and the plastic deformation only has a deviatoric component. At the point of
maximum load, the correlation is moderately high, since $|\vec{E}_0|$ will be a
combination of both plastic strain, and elastic strain due to the stresses.
Once the load is removed, the correlation is very high, since the $|\vec{E}_0|$
is almost entirely determined in terms of plastic strain. At both timepoints,
the volumetric strain is closely correlated with pressure, since there is no
volumetric plastic strain. The volumetric strain at $t=8\times10^5\tsca$ is due
to the residual pressure in the bar.

Since the majority of the load is applied horizontally, the amount that the bar
stretches can be compared to an analytic estimate based on a uniaxial extension
test. Let $\Omega$ be the region where the load is applied. The total
horizontal force per unit length is
\begin{eqnarray}
\bar{F}_x(t) &=&  \int_\Omega 12 \psi F_R(|\vx-\vx_F|) F_T(t) d^2 \vx = 12\psi F_T(t) 2\pi \int_0^{r_F} F_R(r) r \, dr \nonumber \\
&=& 24\psi \pi \left(\frac{r_F^2 (\pi^2-4)}{2\pi^2}\right) F_T(t) = \frac{12\psi r_F^2 (\pi^2 -4)}{\pi} F_T(t) \nonumber \\
&=& 6.48 \times 10^{-6} F_T(t) \sY L  = 55.1 F_T(t) \text{~N/m}
\end{eqnarray}
and hence the maximum load at $t=4\times10^5\tsca$ is $2.59\sY L$ or
$22.0\text{~MN/m}$.

\begin{figure}
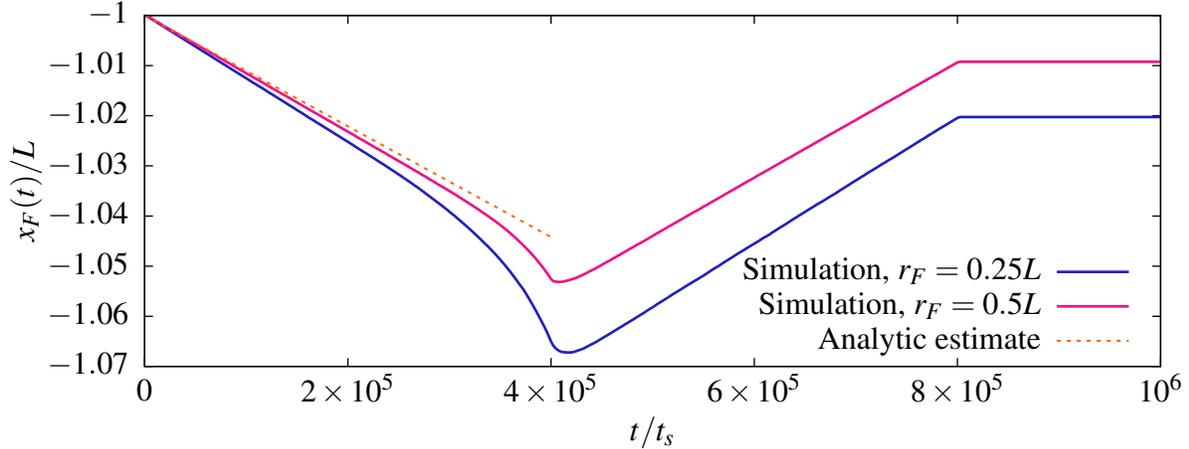

  \include{bd1_ext}
  \caption{Time evolution of the horizontal loading position $x_F(t)$ for two
  stretched bar simulations, compared to an analytic estimate based on a
  uniaxial tension test.\label{fig:bd1_ext}}
\end{figure}

In the plane strain configuration, the effective Young's modulus is given by
$E'=E/(1-\nu^2)$. The loading point $\vx_F(t)$ is initially $3L$ from the fixed
wall and the bar has width $1.3L$. Hence the expected extension as a function
of time is
\begin{equation}
\Delta x_F(t) = \frac{\bar{F}_x(t) 3L}{1.3L E'} = F_T(t) 1.10\times 10^{-7} L.
\end{equation}
Figure~\ref{fig:bd1_ext} shows a plot of the horizontal loading position over
time in the simulation, compared to this analytic estimate. The two curves are
in reasonable agreement, although the gradient of the curve close to $t=0$ has
a slightly larger magnitude in the simulation. This is expected, since
in the simulation the load is localized in a small central region of the bar,
rather than being spread across the whole bar. This is confirmed by the plots
of $|\bsig_0|$ in Fig.~\ref{fig:bdry1_devtem}, which show relatively low levels of
stress at the edges of the bar over the range $-L<x<-0.5L$. To confirm that
this is the source of the discrepancy, a second simulation was carried out
where the diameter of the loading region was doubled to $r_F=0.5L$ while
keeping the total load the same. As expected, the extension in this simulation
is in closer agreement with the analytic estimate.

The plastic deformation of the bar is also evident in Fig.~\ref{fig:bd1_ext}.
As $t$ approaches $4\times 10^5 \tsca$, the rate extension of the loading point
noticeably increases. After the load is removed at $t=8\times 10^5\tsca$, the
loading point does not fully return to its original position. Both simulation
curves show the same trends although less plastic deformation is evident in the
curve for $r_F=0.5L$, since by spreading out the load, and hence stress,
smaller regions of the bar will deform plastically.

\subsection{Transition from the quasi-static simulation to the explicit simulation}
\label{sub:eqtrans}
Since the explicit and quasi-static timestepping methods make use of the same
grids and fields, they can be intermixed, making it possible to simulate
processes with disparate time scales. In a recent paper~\cite{rycroft12b}, we
considered one such situation of dynamic crack propagation, where a bulk
metallic glass was loaded on a time scale of seconds and first accumulates
rather slow plastic deformation, but then fractures on a time scale of
nanoseconds. Here, we consider another case, where a bar is loaded on a
quasi-static timescale and then the load is instantaneously released, making
the bar rapidly oscillate. The simulation domain is $|x| \le0.5L, |y| \le L$
using a $512\times 1024$ grid. The level set function is initialized to be a
vertical bar in the region $|x|<0.25L$ with four holes of radius $0.15L$ at
$x=\pm0.8L,\pm0.4L$ and $y=0$. The bar is attached to the top and bottom
boundaries, and the initial effective temperature is 620~K. The loading
position $\vx_F$ is initially located at the origin. The temporal and spatial
dependencies of the force are given by
\begin{equation}
F_T(t) = \left\{
\begin{array}{ll}
  \frac{t}{\tsca} & \quad \text{for $0\le t \le \trel$,} \\
  0 & \quad \text{for $\trel < \tsca \le \trel+10\tsca$,}
\end{array}
\right.
\qquad
F_R(r) = \left\{
\begin{array}{ll}
  1 + \cos \frac{\pi r}{r_F} & \quad \text{for $r<r_F$,} \\
  0 & \quad \text{for $r\ge r_F$,}
\end{array}
\right.
\label{eq:load2}
\end{equation}
where $\trel=5\times10^5\tsca$ and $r_F=0.15L$. The total force is then given by
\begin{equation}
\vec{F}(\vx,t) = \left(
\begin{array}{c}
  -\psi F_R(|\vx-\vx_F|) F_T(t) \\ 0
\end{array}
\right),
\label{eq:load3}
\end{equation}
where $\psi=5\times10^{-6} \sY/L$. The simulation first uses quasi-static
timesteps of size $625\tsca$ to simulate the time interval $0\le t \le \trel$.
At $t=\trel$, the load reaches its maximum value of $0.105\sY/L$, which is
$0.893\text{~MN/m}$ in the nominal physical units. When the load is removed,
the simulation switches over to explicit timesteps to simulate up to
$t=\trel+10\tsca$.

\begin{figure}
  \begin{center}
    \setlength{\unitlength}{0.0125bp}
    \begin{picture}(36000,40500)(0,0)
      \put(-1800,19600){\parbox{15cm}{\scriptsize \include{bdry2a}}}
      \setlength{\unitlength}{0.0125bp}
      \put(33000,11090){
	\begin{picture}(2000,21100)(0,0)
	  \footnotesize
	  \put(0,1600){\includegraphics[width=12.5pt,height=226pt]{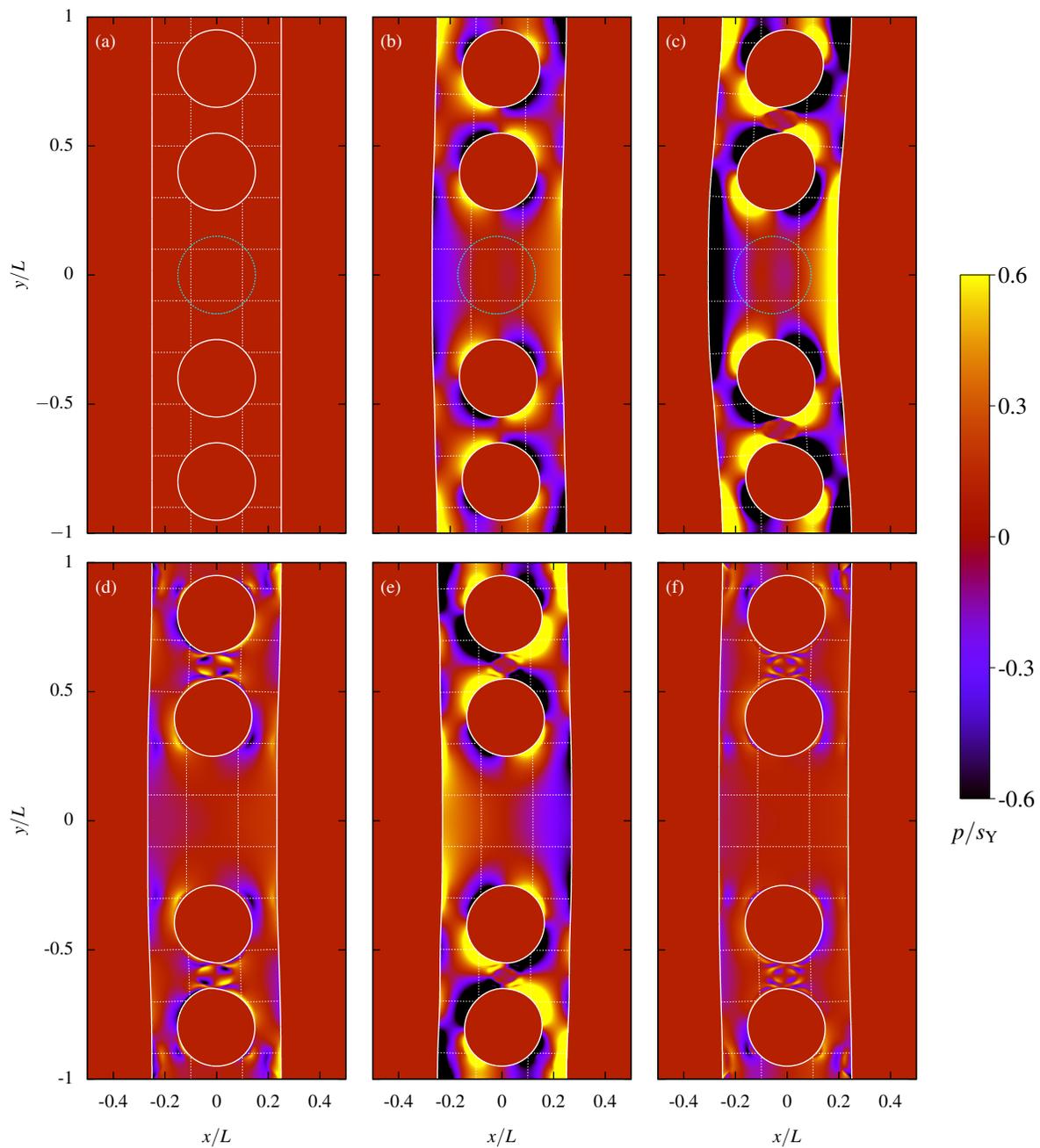}}
	  \put(0,1600){\line(0,1){18000}}
	  \put(0,1600){\line(1,0){1200}}
	  \put(1000,1600){\line(0,1){18000}}
	  \put(0,19600){\line(1,0){1200}}
	  \put(500,400){\makebox(0,0)[c]{$p/\sY$}}
	  \put(1000,6100){\line(1,0){200}}
	  \put(1000,10600){\line(1,0){200}}
	  \put(1000,15100){\line(1,0){200}}
	  \put(1300,1600){\makebox(0,0)[l]{-0.6}}
	  \put(1300,6100){\makebox(0,0)[l]{-0.3}}
	  \put(1300,10600){\makebox(0,0)[l]{0}}
	  \put(1300,15100){\makebox(0,0)[l]{0.3}}
	  \put(1300,19600){\makebox(0,0)[l]{0.6}}
	\end{picture}}
    \end{picture}
  \end{center}
  \caption{Plots of pressure $p$ for the load--release simulation. The top
  three snapshots are of the loading process simulated with the quasi-static
  method, at times of (a) $t=0$, (b) $t=2.5\times10^5\tsca$, and (c)
  $t=\trel=5\times10^5\tsca$. The bottom three snapshots show the dynamics of
  the bar after the load is instantaneously removed, simulated with the explicit
  method, at times of (d) $t=\trel + 2.5\tsca$, (e) $t=\trel +5\tsca$, and (f)
  $t=\trel +7.5\tsca$. The boundary of the bar is shown as the solid white line
  obtained as the zero contour of level set function $\phi$. The thin dashed
  white lines are the contours of the components of the reference map $\bxi$
  and show how the material is deformed. For plots (a) to (c), the dashed cyan
  circle shows the region where the bar is loaded.\label{fig:bdry2a}}
\end{figure}

\begin{figure}
  \begin{center}
    \setlength{\unitlength}{0.0125bp}
    \begin{picture}(36000,40500)(0,0)
      \put(-1800,19600){\parbox{15cm}{\scriptsize \include{bdry2b}}}
      \setlength{\unitlength}{0.0125bp}
      \put(33000,11090){
	\begin{picture}(2000,21100)(0,0)
	  \footnotesize
	  \put(0,1600){\includegraphics[width=12.5pt,height=226pt]{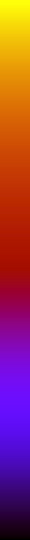}}
	  \put(0,1600){\line(0,1){18000}}
	  \put(0,1600){\line(1,0){1200}}
	  \put(1000,1600){\line(0,1){18000}}
	  \put(0,19600){\line(1,0){1200}}
	  \put(500,400){\makebox(0,0)[c]{$p/\sY$}}
	  \put(1000,6100){\line(1,0){200}}
	  \put(1000,10600){\line(1,0){200}}
	  \put(1000,15100){\line(1,0){200}}
	  \put(1300,1600){\makebox(0,0)[l]{0}}
	  \put(1300,6100){\makebox(0,0)[l]{0.3}}
	  \put(1300,10600){\makebox(0,0)[l]{0.6}}
	  \put(1300,15100){\makebox(0,0)[l]{0.9}}
	  \put(1300,19600){\makebox(0,0)[l]{1.2}}
	\end{picture}}
    \end{picture}
  \end{center}
  \caption{Plots of deviatoric stress $|\bsig_0|$ for the load--release
  simulation. The top three snapshots are of the loading process simulated with
  the quasi-static method, at times of (a) $t=0$, (b) $t=2.5\times10^5\tsca$,
  and (c) $t=\trel=5\times10^5\tsca$. The bottom three snapshots show the
  dynamics of the bar after the load is instantaneously removed, simulated with
  the explicit method, at times of (d) $t=\trel + 2.5\tsca$, (e) $t=\trel
  +5\tsca$, and (f) $t=\trel +7.5\tsca$. The boundary of the bar is shown as
  the solid white line obtained as the zero contour of level set function
  $\phi$. The thin dashed white lines are the contours of the components of the
  reference map $\bxi$ and show how the material is deformed. For plots (a) to
  (c), the dashed cyan circle shows the region where the bar is
  loaded.\label{fig:bdry2b}}
\end{figure}

Figures~\ref{fig:bdry2a} and \ref{fig:bdry2b} show snapshots of the pressure
and deviatoric stress respectively for this simulation. In both figures, the
top row shows snapshots during the quasi-static loading process. As expected,
in the middle of the bar, negative pressures grow on the left of the bar as
it is stretched, and positive pressures grow on the right of the bar as it
is compressed. The largest deviatoric stresses develop in the small regions
between each pair of holes at $\vx=(0,\pm\frac{L}{2})$, and also at edges of
the bar close to the top and bottom boundaries. In both of these regions,
$|\bsig_0|$ exceeds $\sY$ and hence plastic deformation takes place.

In Figs.~\ref{fig:bdry2a} and \ref{fig:bdry2b} the bottom row of snapshots show
the bar at several points after the load has been released. As soon as the load
is released, elastic waves rapidly propagate through the bar, and the stress
imbalance pushes the bar rightward. Figs.~\ref{fig:bdry2a}(d) and
\ref{fig:bdry2b}(d) show the bar when it first reaches an approximately
vertical state. Some small concentrations of pressures and deviatoric stress
are visible in the regions that underwent plastic deformation. In
Fig.~\ref{fig:bdry2a}(e) and \ref{fig:bdry2b}(e), the bar is shown at its
maximal rightward extent. Very large deviatoric stresses are visible in the
regions between each pair of holes. After this timepoint, the bar begins to
move leftward. Fig.~\ref{fig:bdry2a}(f) and \ref{fig:bdry2b}(f) show the bar
when it becomes vertical for the second time.

\begin{figure}
    {\footnotesize
    \input{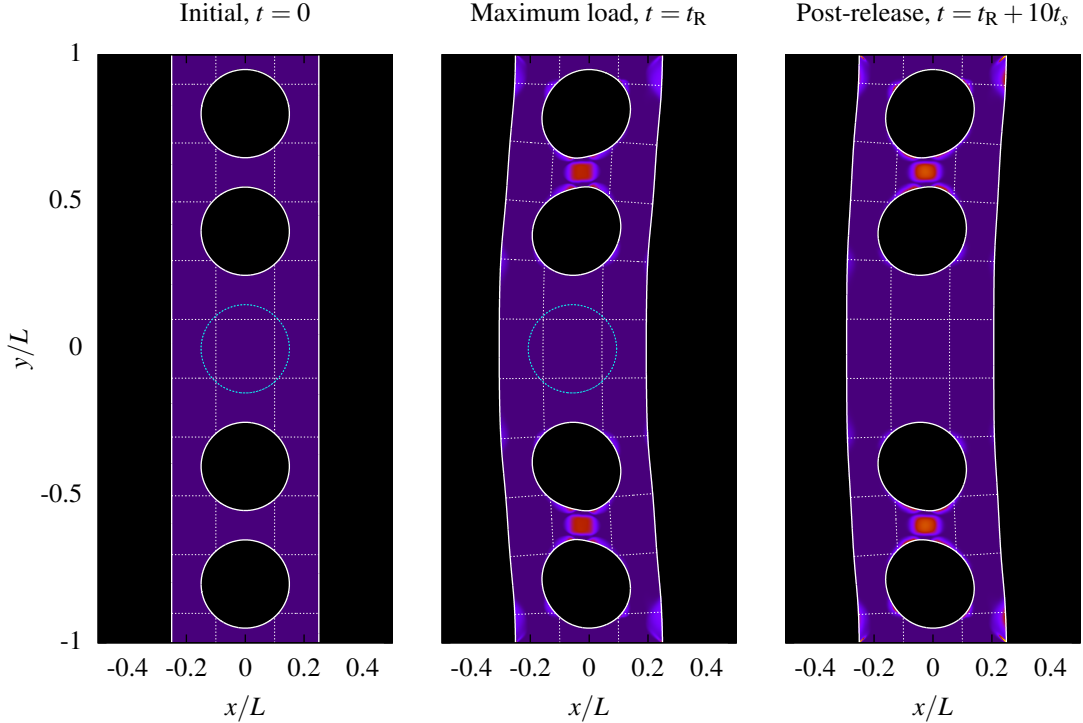}}
  \caption{Plots of the effective temperature $\chi$ for the load--release
  simulation at three time points. The boundary of the bar is shown as the
  solid white line obtained as the zero contour of level set function $\phi$.
  The thin dashed white lines are the contours of the components of the
  reference map $\bxi$ and show how the material is deformed. For plots (a) and
  (b), the dashed cyan circle shows the region where the bar is loaded. The
  color gradient for the effective temperature is the same as that used in
  Fig.~\ref{fig:basic_tem}.\label{fig:bdry2c}}
\end{figure}

Figure~\ref{fig:bdry2c} shows the effective temperature in this simulation at
three time points. At $t=\trel$, as expected, an increase effective temperature
is visible in the regions between the holes, and near the top and bottom
boundaries. At $t=\trel+10\tsca$, after the bar has undergone the rapid
oscillatory motion, further increases in $\chi$ are visible in the regions
between the holes. Because the oscillatory motion creates large deviatoric
stresses up to $1.9\sY$, and the plasticity model specified in
Eqs.~\ref{eq:stzdpl} and \ref{eq:stzceq} has an exponential dependence on
$|\bsig_0|$, noticeable plastic deformation can occur on a very short
timescale. This is a consequence of the simplified choice of the plasticity
model discussed in Subsec.~\ref{sub:plasticity}.

The loading phase and release phase differ by more than four orders of
magnitude in duration, and this example therefore highlights the ability to
simulate phenomena across a wide range of timescales. It may also be possible
to carry out an opposite transition from an explicit simulation to a
quasi-static simulation, although this would require careful consideration of
any elastic waves in the explicit simulation, which would immediately disappear
after a single quasi-static projection step. In the free boundary examples
presented here and in the previous subsection, it has been possible to
determine {\it a priori} whether the quasi-static method or the explicit method
should be applied, but this may not be the case in general, particularly since
in an elastoplastic material the relevant timescales may dynamically change. In
Fig.~\ref{fig:bd1_ext}, the loading position starts to move more quickly near
the time of maximum load, due to the positive feedback between effective
temperature and $\Dpl$, and for larger loads, the motion may become so great
that quasi-staticity may no longer be a reasonable assumption. We expect that
this can be quantified by examining the size of the projection required to
restore quasi-staticity, creating the possibility of automatically selecting
the correct time-integration method to use, although we leave this for the
subject of future work.

\section{Conclusion}
Building on a mathematical correspondence with the incompressible
Navier--Stokes equations, we have developed a numerical method for simulating
the deformation of elastoplastic materials in the quasi-static limit that is
analogous to the projection method in fluid mechanics~\cite{chorin68}. The new
method is most suitable for materials that can be well-described by the
additive decomposition of the deformation rate into elastic and plastic parts.
It is well-suited for a large class of materials (\textit{e.g.} metals and
amorphous solids such as metallic glasses), which typically undergo small
elastic deformations and feature large elastic wave speeds, making many plastic
deformation problems intrinsically quasi-static. In such situations, the new
method allows simulating realistic loading rates, which would be prohibitively
computationally expensive using explicit methods.

The method is naturally implemented in an Eulerian framework. It is
particularly well-suited to cases of straightforward boundary conditions, such
as the simple shear experiments discussed in Section~\ref{sec:shearing}. We
examined several basic features of shear band development in the STZ plasticity
model, but the method could be adapted to look at a wide variety of problems in
elastoplasticity, using STZ plasticity or other plasticity models. For example,
detailed questions of shear band nucleation and growth, shear band interaction,
or the role of structural inhomogeneities can be examined, and will be
addressed elsewhere. Models with more complex physics, such as a coupling to
real temperature evolving according to the diffusion equation, are also
straightforward to incorporate. The derivation of the method should also be
generalizable to the case of a non-constant and anisotropic stiffness tensor
$\tC$, and other objective stress rates, such as the Truesdell or Oldroyd
stress rates.

As described in Section~\ref{sec:free}, the method can also be applied to
problems involving moving free boundaries by using a suitable description of
the boundary, such as the level set method. This framework may be well-suited
to various fluid--structure interaction problems, offering some of the same
advantages as the Eulerian hyperelasticity
methods~\cite{plohr88,trangenstein91,liu01,cottet08,kamrin12}. It may be
interesting to examine the case of a quasi-static elastoplastic material
interacting with an incompressible fluid, which would require a double
projection to enforce both fluid incompressibility and solid quasi-staticity. As
demonstrated in the final example in Subsec.~\ref{sub:eqtrans}, the method can
also be intermixed with explicit timestepping, making it possible to simulate
phenomena on multiple timescales.

The method presented here is underpinned by a close mathematical connection
between the variables $(p,\vv)$ in the incompressible Navier--Stokes equations
and $(\vv,\bsig)$ in quasi-static elastoplasticity. This connection may
therefore allow mathematical results for fluid mechanics to be translated to
elastoplasticity. The incompressible limit of fluid mechanics has been
extensively analyzed, often by examining the limit of small Mach number $M$,
describing the ratio of a typical velocity to the sound speed, and playing a
similar role to the artificial compressibility parameter~\cite{chorin67}.
Klainerman and Majda established that the solutions of the incompressible
Navier--Stokes equations will match the compressible Navier--Stokes equations
in the limit of small $M$~\cite{klainerman82}. In the context of turbulent
combustion, where the Navier--Stokes equations are coupled to a
reaction--diffusion equation, the zero Mach number limit has been examined by
introducing perturbative expansions of the fields in powers of
$M$~\cite{majda85,embid87}. These mathematical approaches provide possible
avenues to establish rigorously that, in the long timescale limit, solutions to
the full elastoplastic system will match the elastoplastic system with the
quasi-staticity constraint.

A variety of advanced numerical approaches based on the fluid projection method
have been developed, and it may be possible to translate these to
elastoplasticity. Currently, the projection step that we employ is first-order
accurate, but it is likely that high-order fluid projection
methods~\cite{bell89,bell92,puckett97,brown01} could be adapted to the
elastoplastic framework. The fluid projection step has also been implemented
using finite elements within a finite-difference
calculation~\cite{almgren96,yu03}, which has the advantage of simplifying
boundary conditions, and this may provide a simpler solution for quasi-static
elastoplasticity than the extrapolation formulae introduced in
Subsec.~\ref{sub:quasi_boundary}. The fluid projection method has also been
implemented on adaptive resolution grids~\cite{min06}, and if this was applied
to elastoplasticity, it would allow for the investigation of the detailed
structure of the localized shear bands that are a common feature of plasticity
models. All of these interesting possibilities and directions should be
systematically explored in future investigations.

\section*{Acknowledgments}
The authors thank Prof.~Alexandre J.~Chorin (University of California,
Berkeley), Prof.~Ken Kamrin (Massachusetts Institute of Technology),
Prof.~James. R.~Rice (Harvard University), and Dr.~Manish Vasoya (Weizmann
Institute of Science) for useful discussions about this work. C.~H.~Rycroft was
supported by the National Science Foundation under Grant No.~DMR-1409560, and by
the Director, Office of Science, Computational and Technology Research,
U.S.~Department of Energy under contract number DE-AC02-05CH11231.
E.~Bouchbinder acknowledges support from the Minerva Foundation with funding
from the Federal German Ministry for Education and Research, the Israel Science
Foundation (Grant No.~712/12), the Harold Perlman Family Foundation, and the
William Z.~and Eda Bess Novick Young Scientist Fund.

\section*{Author contributions}
C.~H.~Rycroft developed the mathematical connection to the fluid projection
method, formulated and programmed the explicit and quasi-static simulations,
analyzed the results, created the figures, and wrote the manuscript. Y.~Sui
carried out initial tests of the explicit simulation method in the shear
banding geometry of Section \ref{sec:shearing}. E.~Bouchbinder identified the
problem of simulating hypo-elastoplasticity in the quasi-static regime,
identified the physical importance of simulating realistic loading rates,
introduced the constitutive relations, and was involved in writing the
manuscript.

\appendix
\section{Uniqueness of solution to the algebraic problem in Eq.~\ref{eq:sproj3}}
\label{app:unique}
In the quasi-static projection method, it is necessary to solve the algebraic
problem given in Eq.~\ref{eq:sproj3}, which can be rewritten as
\begin{equation}
  \nabla \cdot \bsig_* = -\Delta t \nabla \cdot (\tC : \nabla \vv_{n+1})
\end{equation}
by taking into account the symmetries of $\tC$. Suppose that this equation must
be solved on a fixed domain $\Omega$ where Dirichlet conditions for velocity
are prescribed on the boundary $\p \Omega$. Suppose that a second solution
$\vv^\prime_{n+1}$ exists. Hence the function $\vec{w} = \vv^\prime_{n+1} -
\vv_{n+1}$ satisfies
\begin{equation}
  0 = \nabla \cdot (\tC : \nabla \vec{w})
  \label{eq:projdiff}
\end{equation}
in $\Omega$, and $\vec{w}=\vec{0}$ on $\p \Omega$. Multiplying the right hand
side of Eq.~\ref{eq:projdiff} by $\vec{w}$ and integrating gives
\begin{equation}
  0 = \int_\Omega \vec{w} \cdot ( \nabla \cdot (\tC : \nabla \vec{w})) d^3\vx =
  \int_{\p \Omega} \nor \cdot ( \vec{w} \cdot (\tC : \nabla \vec{w} )) dS - \int_\Omega (\nabla \vec{w}) : (\tC : (\nabla \vec{w})) d^3\vx.
  \label{eq:uniqueint}
\end{equation}
The boundary integral will vanish since $\vec{w}=\vec{0}$ there, and hence
\begin{equation}
  0 = \int_\Omega (\nabla \vec{w}) : \tC : (\nabla \vec{w}) d^3\vx.
\end{equation}
Since $\tC$ is positive definite, it follows that $\nabla \vec{w}=\ten{0}$
and therefore $\vec{w}$ is a constant. Assuming $\p \Omega\ne \emptyset$,
the boundary condition will enforce that $\vec{w}=\vec{0}$, and hence that
$\vv^\prime_{n+1} = \vv_{n+1}$ so that Eq.~\ref{eq:sproj3} has a unique
solution.

The above argument will also apply for traction-free boundary conditions
discussed in Section~\ref{sec:free}. Equation~\ref{eq:s_nbc} will lead to a
Neumann-like condition $\nor \cdot (\tC : \nabla \vec{w})=\vec{0}$, which will
also lead to the boundary term in Eq.~\ref{eq:uniqueint} vanishing. In the case
when only traction-free boundary conditions are applied, $\vec{w}$ will be a
constant, so that $\vv^\prime_{n+1}$ and $\vv_{n+1}$ are equal up to a
constant, which as discussed in Section~\ref{sec:free} is physically
reasonable.

\section{Adaptive sub-stepping}
\label{app:adapt}
As described in Subsec.~\ref{sub:directsim}, the plastic deformation $\dpl$ grows
rapidly when $\bs>\sY$, and this can cause the forward Euler timestepping
procedure to lose accuracy, so that in a single timestep of size $\Dt$, the
change $\Delta\bs$ in the deviatoric stress may be very large and substantially
overshoot the yield surface. To solve this, an adaptive timestepping procedure
is used that considers the coupled system of $\bs$ and $\chi$ over the timestep
$\Delta t$, in isolation from other terms. The procedure divides the interval
$\Dt$ into a number of substeps so that the change $\Delta \bs$ at each substep
remains within a fixed tolerance $\afac$; throughout this study, a value of
$\afac=0.2\%$ is used. To begin, the values of deviatoric stress and effective
temperature at a given gridpoint are stored as $\bs_0$ and $\chi_0$
respectively. The following algorithm is then used:
\begin{algorithmic}
  \STATE $\alpha=0$
  \STATE $t_\text{R}= \Dt$
  \STATE $Q=\text{true}$
  \WHILE{$Q$}
    \STATE Evaluate $D'= 2\mu\dpl(\bs_\alpha,\chi_\alpha)/\sY$
    \STATE Evaluate $F = F(\bs_\alpha,\chi_\alpha)$
    \IF{$D' t_\text{R} >\afac$}
      \STATE $t_\text{S} \leftarrow \afac/D'$
      \STATE $t_\text{R} \leftarrow t_\text{R} -t_\text{S}$
    \ELSE
      \STATE $t_\text{S} \leftarrow t_\text{R}$
      \STATE $Q \leftarrow \text{false}$
    \ENDIF
    \STATE $\bs_{\alpha+1} \leftarrow \bs_\alpha - t_\text{S} D' \sY$
    \STATE $\chi_{\alpha+1} \leftarrow \chi_\alpha + t_\text{S} F$
    \STATE $\alpha \leftarrow \alpha+1$
  \ENDWHILE
\end{algorithmic}
Here, a left arrow is used to signify a variable being updated, and the functions
$\dpl$ and $F$ are derived from Eqs.~\ref{eq:stzdpl} and \ref{eq:chiup},
respectively. Within the algorithm, the variable $t_\text{R}$ holds the
remaining portion of the time interval to be considered. In the main loop, the
algorithm determines whether the value of $\Delta\bs$ for a timestep of size
$t_\text{R}$ is within the threshold $\eta$. If so, the algorithm takes a
timestep of size $t_\text{R}$ and terminates. Otherwise, it steps forward by
the time interval $t_\text{S}$ that makes $\Delta \bs$ exactly match the
threshold; it then subtracts $t_\text{S}$ from $t_\text{R}$ and repeats. Once
the algorithm has terminated, corrected versions of plastic deformation and
effective temperature change are evaluated according to
\begin{equation}
  \label{eq:adapt_cor}
\tilde{D}_\text{pl} = \frac{\bs_0-\bs_\alpha}{2\mu \,\Dt}, \qquad \tilde{F} = \frac{\chi_\alpha-\chi_0}{\Dt}.
\end{equation}
These values are then used within the main finite-difference updates given in
Eqs.~\ref{eq:direct_sstart} to \ref{eq:direct_send} for the explicit simulation
and Eqs.~\ref{eq:qs_sstart} to \ref{eq:qs_send} for the quasi-static
simulation. If $\eta$ is sufficiently large or $\Delta t$ is sufficiently
small, so that the algorithm always terminates after a single step, then the
main finite-difference updates reduce to the standard, fixed-timestep forward
Euler procedure.

\bibliography{elas}


\end{document}